\documentclass{article}

\usepackage{amsmath}
\usepackage{amssymb,amsthm}
\usepackage{latexsym}
\usepackage{chemarr}
  \usepackage{paralist}
  \usepackage[pdftex]{graphics} 
  \usepackage{epsfig} 

\pdfoutput=1

\newtheorem{theorem}{Theorem}[section]

\newtheorem{lemma}[theorem]{Lemma}

\newtheorem{assumption}{Assumption}

\theoremstyle{definition}

\newtheorem{remark}{Remark}

\newtheorem{axiom}{Axiom}
\newtheorem{defin}{Definition}
\newtheorem{theo}{Theorem}[section]
\newtheorem{propo}{Proposition}
\newtheorem{corol}{Corollary}
\newcommand{\bax}{\begin{axiom}}
\newcommand{\eax}{\end{axiom}}
\newcommand{\bass}{\begin{assumption}}
\newcommand{\eass}{\end{assumption}}
\newcommand{\bdefi}{\begin{defin}}
\newcommand{\edefi}{\end{defin}}
\newcommand{\bth}{\begin{theo}}
\renewcommand{\eth}{\end{theo}}
\newcommand{\bprop}{\begin{propo}}
\newcommand{\eprop}{\end{propo}}
\newcommand{\blem}{\begin{lemma}}
\newcommand{\elem}{\end{lemma}}
\newcommand{\bcor}{\begin{corol}}
\newcommand{\ecor}{\end{corol}}
\newcommand{\brem}{\begin{remark}}
\newcommand{\erem}{\end{remark}}
\newcommand{\bpf}{\begin{proof}}
\newcommand{\epf}{\end{proof}}

\newcommand{\noi}{\noindent}

\newcommand{\bn}{{\bf n}}
\newcommand{\vn}{\vec{\bf n}}
\newcommand{\bk}{{\bf k}}

\newcommand{\ve}{\vec{\bf e}}
\newcommand{\bx}{{\bf x}}
\newcommand{\vx}{\vec{\bf x}}

\newcommand{\vxi}{\vec{\xi}}
\newcommand{\vmu}{{\vec{\mu}}}
\newcommand{\field}[1]{\mathbb{#1}}

\newcommand{\R}{\field{R}}

\newcommand{\Z}{\field{Z}}
\newcommand{\N}{\field{N}}
\newcommand{\X}{\field{X}}
\newcommand{\LL}{\field{L}}
\newcommand{\pa}{\partial}
\newcommand{\supp}{\mbox{supp}}

\newcommand{\ra}{\rightarrow}

\newcommand{\rlha}{\xrightleftharpoons}
\newcommand{\de}{\delta}

\newcommand{\ga}{\gamma}
\newcommand{\eps}{\epsilon}

\newcommand{\I}{{\mathcal I}}
\newcommand{\vI}{\vec{\mathcal I}}

\newcommand{\cX}{{\mathcal X}}
\newcommand{\vcX}{{\vec{\mathcal X}}}

\newcommand{\Eop}{{\bf E}}
\newcommand{\Dop}{{\bf \Delta}}
\newcommand{\Lop}{{\widehat{{\mathcal L}}}}
\newcommand{\Kop}{{\widehat{{\mathcal K}}}}
\newcommand{\Ld}{{{{\mathcal L}}}}
\newcommand{\Kd}{{{{\mathcal K}}}}
\newcommand{\Pro}{{{{\mathcal P}}}}

\newcommand{\id}{{\bf id}}
\newcommand{\bPi}{{\bf \Pi}}

\newcommand{\be}{{\bf 1}}
\newcommand{\bI}{{\bf I}}
\newcommand{\Sig}{\Sigma}
\newcommand{\sig}{\sigma}
\newcommand{\vde}{\vec{\delta}}
\newcommand{\der}{\textrm{d}}

\newcommand{\app}{\approx}

\title{Multiscale Analysis of Reaction Networks}

\author{L.Sbano\footnote{\textbf{sbano@maths.warwick.ac.uk}} and M.Kirkilionis\footnote{\textbf{ mak@maths.warwick.ac.uk}}\\ Mathematics Institute, University of Warwick, CV4 7AL Coventry, UK}

\begin{document}
\maketitle



   %



\begin{abstract}
In most natural sciences there is currently the insight  that it is necessary to bridge gaps between different processes which can be observed on different scales. This is especially true in the field of chemical reactions where the abilities to form bonds between different types of atoms  and molecules create much of the properties we experience in our everyday life, especially in all biological activity. There are essentially two types of  processes related to biochemical reaction networks, the interactions among molecules and  interactions involving their conformational changes, so in a sense, their internal state. The first type of processes can be conveniently approximated by the so-called mass-action kinetics, but this is not necessarily so for the second kind where molecular states do not define any kind of density or concentration.  In this paper we demonstrate the necessity to study reaction networks in a stochastic formulation for which we can construct a coherent approximation in terms of specific space-time scales and the number of particles. The continuum limit procedure naturally creates 
 equations of Fokker-Planck type where the evolution of the concentration occurs on a slower time scale when compared to the evolution of the conformational changes, for example triggered by binding or unbinding events with other (typically smaller) molecules. We apply the asymptotic theory to derive the effective, i.e. macroscopic dynamics of the biochemical reaction system. The theory can also be applied to other processes where entities can be described by finitely many internal states, with changes of states occuring by arrival of other entities described by a birth-death process.
 \end{abstract}

\section{Introduction}
Systems formed by a large number of biochemical reactions are often considered paramount examples of complex systems. Such systems are formed by a set of interactions among various species of molecules forming new, larger species. Moreover there are interactions involving conformational changes coinciding with binding/unbinding events of typically smaller molecules.
It is important to note that the description of interactions depends on the choice 
of the scales at which the entire system is analysed. The microscopic description of a reaction system is usually fairly well understood in its general features. At atomic scale the necessary theory is provided by Quantum Mechanics, at molecular level there are different types of kinetic theories. After a heuristic up-scaling most reaction systems can be sufficiently well described by mass-action kinetics, which is a mean-field approximation of the fully stochastic description \cite{Kurtz3},\cite{Hanggi}.\\ 

We now consider systems at scales to be considered mesoscopic. These are precisely the scales of any kinetic theory.  Let us denote with $\vde$ a vector describing the selected  space scale and the number 
of particles in the system, whereas $\tau$ denotes the system time scale. With fixed $\vde,\tau$ we can typically look at reactions specified by the reaction rates $k(\vde,\tau)$, where we incorporated scale dependence. Given these reaction rates it is possible to construct  the associated dynamics in terms of the \emph{master equation} (ME), which is the appropriate probabilistic description of the dynamics precisely at the given scales $\vde,\tau$. If the system contains both interactions among particles and interactions involving conformational changes possibly with additional  binding/unbinding events of smaller molecules, then the ME turns out to be a combination of two types of operators: one describing birth-death processes 
 with   infinite possible states, and the other one governing the evolution in the finite state space. This finite state space (denoted by $\Sigma$)  describes  conformational changes and mutual binding/unbinding of molecules. \\

The process of removing the scales ($\vde\ra 0$ and $\tau\ra 0$) under the condition to keep finite reaction rates $k(\vde,\tau)$ is called \emph{continuum limit}. This process produces a Fokker-Planck equation (FPE) that describes the effective time evolution of the probability distribution of the state of the system. The continuum limit is dependent on fixing a relation among $\vde,\tau$. A better known and typical case is the derivation of a diffusion equation, where $|\vde|^2/\tau=D>0$ is kept finite.  We shall show that the choices involved in the continuum limit determine a FPE where 
the time scale of the evolution of molecular concentrations is larger, i.e. longer than the time scale at which the evolution in 
finite state space $\Sigma$ takes place.  The formulation of the continuum limit will be done following the Trotter approximation method, see \cite{Trotter}, or \cite{Pazy}. We shall illustrate how the limit for $\vde\ra 0$ and $\tau\ra 0$ leads naturally to the use of asymptotic analysis and  an
adiabatic theory to study the FPE. Previous applications of these ideas to  study chemical reaction networks can be found in \cite{GME} and \cite{kepler-elston}. The multi-scale analysis for such systems has been studied extensively, see for example \cite{Kurtz} and \cite{Pavliotis}. In this paper we present the asymptotic solution of the FPE motivated by the continuum limit. We give a general formulation of the approach where  the  stochastic processes involved are not necessarily Markovian. Nevertheless  our main results will deal only with reaction systems involving elementary processes which
are Markovian. In this setting the particles will undergo diffusion and the finite states will evolve 
according to a Markov chain logic. \\

A set of reactions can naturally be  described as a network and more precisely as a  graph. Indeed in this paper we show that graph-theoretic notions can be 
used at the very beginning of modelling as a tool to understand the possible processes. The associated graph is generally called the \emph{Interaction Graph} (IG), 
its vertices are the possible states and its edges correspond to the interactions leading to state switches. 
The IG is then modified throughout the analysis, in fact the continuum limit produces variations in  the vertices and in the edges. In particular 
it turns out that the leading order term of the asymptotic expansion is a deterministic dynamics termed \emph{average dynamics}. 
The average dynamics is determined by a vector field resulting from the average of a finite family of vector fields $\{\cX^{(\sig)}\}_{\sig\in\Sig}$ taken against the invariant measure of the finite Markov chain (MC) on $\Sig$. The IG associated to the average vector field will result as a combination - resembling an average - of the IGs associated to vector fields describing each single finite state. The construction of the average dynamics and its IG can be seen as a first step to connect the stochastic description 
to the classical differential equations approach. To explore the possible applications of graph theory to 
reaction systems given in terms of differential equations the reader could look at the review \cite{Mirela}.This paper deals with different graph theoretic methods giving information on the qualitative behaviour of the reaction system once it has been established on the mesoscopic or macroscopic scale. \\

The continuum limit and the asymptotic analysis will be illustrated by three simple examples: a particle with two internal states diffusing on a line, and two possible schemes for a molecular switch. In these systems we show how to identify  the scaling regimes which characterise the dynamics and  the adiabatic expansion for the associated Fokker-Planck equations. We also show how the network structure 
 of the reactions affects the expansion, in particular with respect to the leading order term, i.e. the \emph{average vector field}  and the appearance of noise.

\section{General formulation}
Let us consider $N$ species $n_1,...,n_N$ of particles each of which can take 
any value in a $N$ dimensional lattice $\LL$ and a variable $\sig$ which can assume values in a fine set $\Sig$ with $|\Sig|=M$. At any time $t$ the system has its configuration determined by 

\[(\vn,\sig)\in\LL\times\Sig.\]

\brem
Note that we did not include explicitly the space variable. This can be easily done by a suitable enlargement of the lattice $\LL$. 
\erem
\noi The time evolution of the system is stochastic and therefore the main object of interest
is the probability measure

\[P(\vn,\sig,t),\mbox{ normalised by }\sum_{\vn\in\LL}\sum_{\sig\in\Sig}P(\vn,\sig,t)=1.\]

\subsubsection*{Dynamical processes}
\noi The time evolution of $P(\bn,\sig,t)$ is determined by certain processes which affect the state of the system. It is 
their nature and characteristics which prescribe the form of the dynamical equations. 
The dynamical processes are strongly related to the scale at which the system is considered. Let us fix $N+1$ scales, i.e.

\begin{itemize}
\item $\tau$, the time scale,
\item a vector $\vde=(\de_1,...,\de_N)$, the natural length scales of the generators of $\LL$.
\end{itemize}

The possible processes we shall consider have a \emph{general diffusive behaviour}, that is each process is characterised by 
having a specific \emph{waiting time probability distribution}, generically denoted by $\psi(t)$. It is important to note that many application will require that $\psi(t)$ is not necessarily exponential, for example in processes generating sub-diffusive behaviour. With fixed $\psi(t)$ and a given process 
we know that the dynamical transition produced by that process will take place in the time interval $[t_1,t_2]$ with a probability given by

\[\int_{t_1}^{t_2}\der t\,\psi(t).\]

\noi The knowledge of the waiting time distribution is in general related to the understanding of the processes and their relevant interactions 
at the scale identified by $\tau$ and $\vde$, therefore it is expected that the functions $\psi$ are dependent on such scales. Upon these observations we can now set up the microscopic reaction schemes linked to the processes. These are classified according to the following list: \vskip4mm

\begin{itemize}
\item[(P1)] $(n_1,...,n_i,...,n_N, \sig)\mapsto (n_1,...,n_i\pm\de_i,...,n_N, \sig)$, with waiting time distribution $\psi_{n_i}(t;\de_i,\tau)$;\\[2mm]
\item[(P2)] $(n_1,...,n_N, \sig)\mapsto (n_1,...,n_N, \sig')$, with waiting time distribution $\psi_{\sig}(t;\tau)$;\\[2mm]
\item[(P3)] $(n_1,...,n_i,...,n_N, \sig)\mapsto (n_1,...,n_i\pm\de_i,...,n_N, \sig')$, with waiting time distribution $\psi_{\sig,n_i}(t;\de_i,\tau)$.\\
\end{itemize}

Here $(P1)$ describes the appearance or annihilation of particles, without changes of  any of the internal states as described by $\sigma$, $(P2)$ describes a transition of the internal states from $\sigma$ to $\sigma^\prime$ while fixing the number of particles in the system, and $(P3)$ describes the simultaneous transition of internal states linked with the appearance or disappearance of a particle of a certain type. Here we must distinguish two cases for the interpretation of  $\LL$. If we model a spatially averaged system we only consider  $\LL$ as representing the species number, so appearance or disappearance models whether particles enter or leave the system. If  $\LL$ includes spatial positions then appearance or disappearance is interpreted with respect to any local position. See also Figure \ref{fig:Lattice-1D}.

\subsubsection*{General Master Equation}
Each process can in principle occur with a specific waiting time governed by its own distribution function. This 
implies  that the evolution is described through 
 a general master equation (GME) (see \cite{GME}). The discrete form of this equation is 

\begin{equation}
\label{discrete_GME}
\begin{split}
P(\vn,\sig,t+\tau)=\sum_{\sig'\in\Sig}\int_0^t\der t'\,\Ld_{\sig\sig'}[\vn,\vde,\tau,t-t']\,P(\vn,\sig',t')+&\\
+\sum_{\sig'\in\Sig}\int_0^t\der t'\Kd_{\sig\sig'}[\vn,\vde,\tau,t-t']P(\vn,\sig',t'),&
\end{split}
\end{equation}

where

\begin{itemize}
\item $\Ld$ is a $M\times M$ matrix whose entries depend on $\vn,\vde,t$ and on 
the waiting time distributions defined in (P1) and (P3). 
In particular it will be useful to write $\Ld$ by means of the operators $\Eop^\pm_i$ defined by
\[\Eop^{\pm}_i f(n_1,...,n_N)=f(n_1,...,n_i\pm\de_i,...,n_N)\mbox{ for any $f:\LL\ra\R$}.\]  

\item $\Kd$ is a $M\times M$ matrix whose entries depend on $\vn,\vde,t$ and on 
waiting time distributions defined in (P2).
\end{itemize}

\brem
The structure of the GME is essentially the one introduced in  \cite{GME} and \cite{Weiss} to describe the 
continuous time random walk (CTRW).  It is worth noting that a standard random walk (RW) 
can always be considered as a special case of a CTRW.
\erem

\noi Note that the normalisation condition for the probability requires

\[\sum_{\vn\in\LL}\,\sum_{\sig',\sig\in\Sig}\int_0^t\der t'\,\left\{\Ld_{\sig\sig'}[\vn,\vde,\tau,t-t']+\Kd_{\sig\sig'}[\vn,\vde,\tau,t-t']\right\}P(\vn,\sig',t)=1\]
for any $P(\vn,\sig,t)$.

\subsubsection*{A graph for the General Master Equation}
The structure of equation (\ref{discrete_GME}) allows a useful interpretation in terms of associated graphs. 
\bdefi
We denote by $\vI(\Ld,\Kd)$ the graph whose vertex set is $V=\LL\times\Sig$ and edge set
$E(\Ld,\Kd)$, where the directed link (arrow) $\ve_{(\vn_i,\sig_i),(\vn_j,\sig_j)}$ is present if both
$\Ld$ and $\Kd$ allow the transition $(\vn_i,\sig_i)\mapsto (\vn_j,\sig_j)$. In general this graph will have loops.
\edefi

\begin{figure}[htbp] 
   \centering
   \includegraphics[scale=0.4]{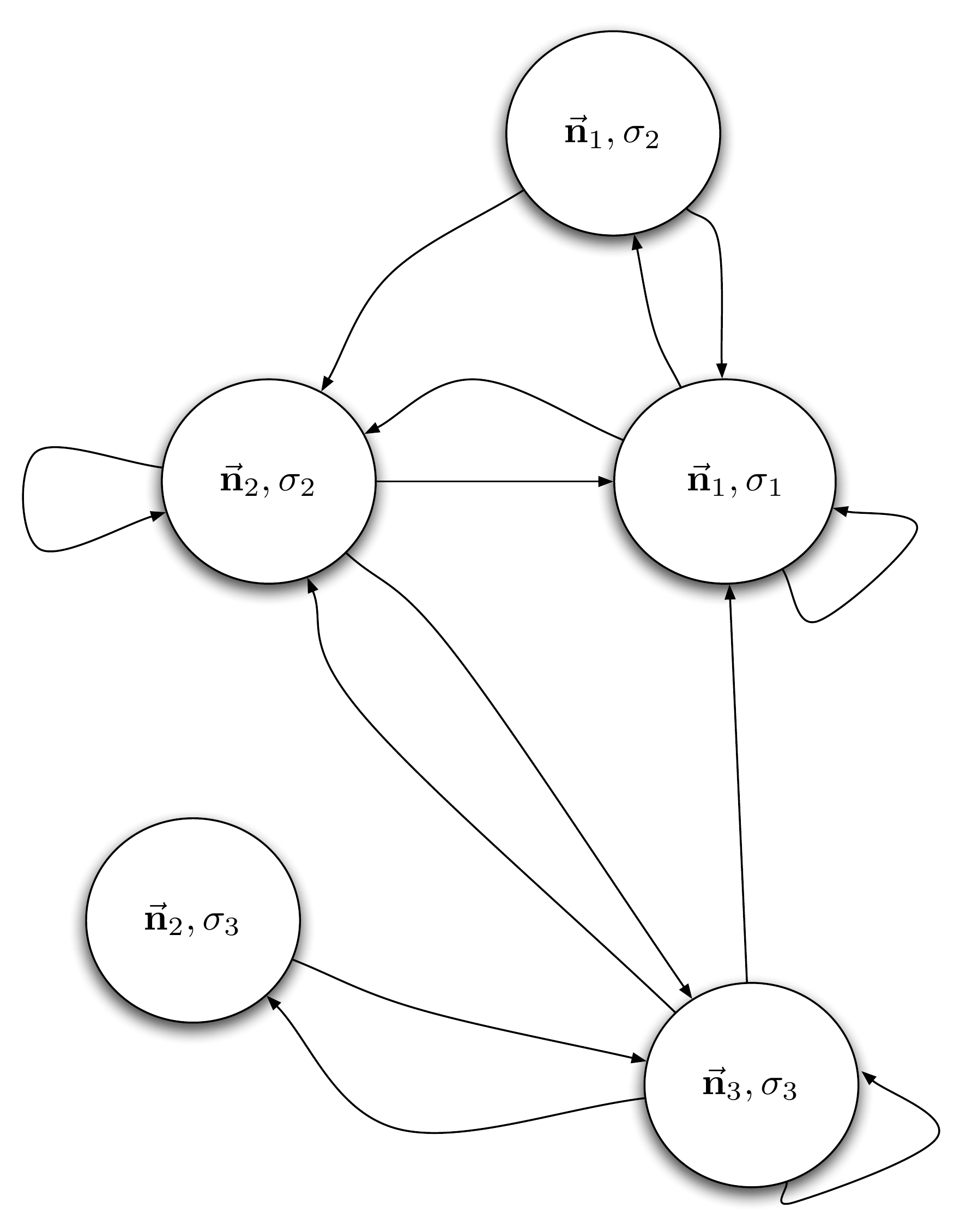} 
   \caption{A small portion of a general graph $\vI(\Ld,\Kd)$}
   \label{fig:GG}
\end{figure}

\noi An illustration of this graph can be seen in Figure \ref{fig:GG}.

\begin{figure}[htbp] 
   \centering
   \includegraphics[scale=0.4]{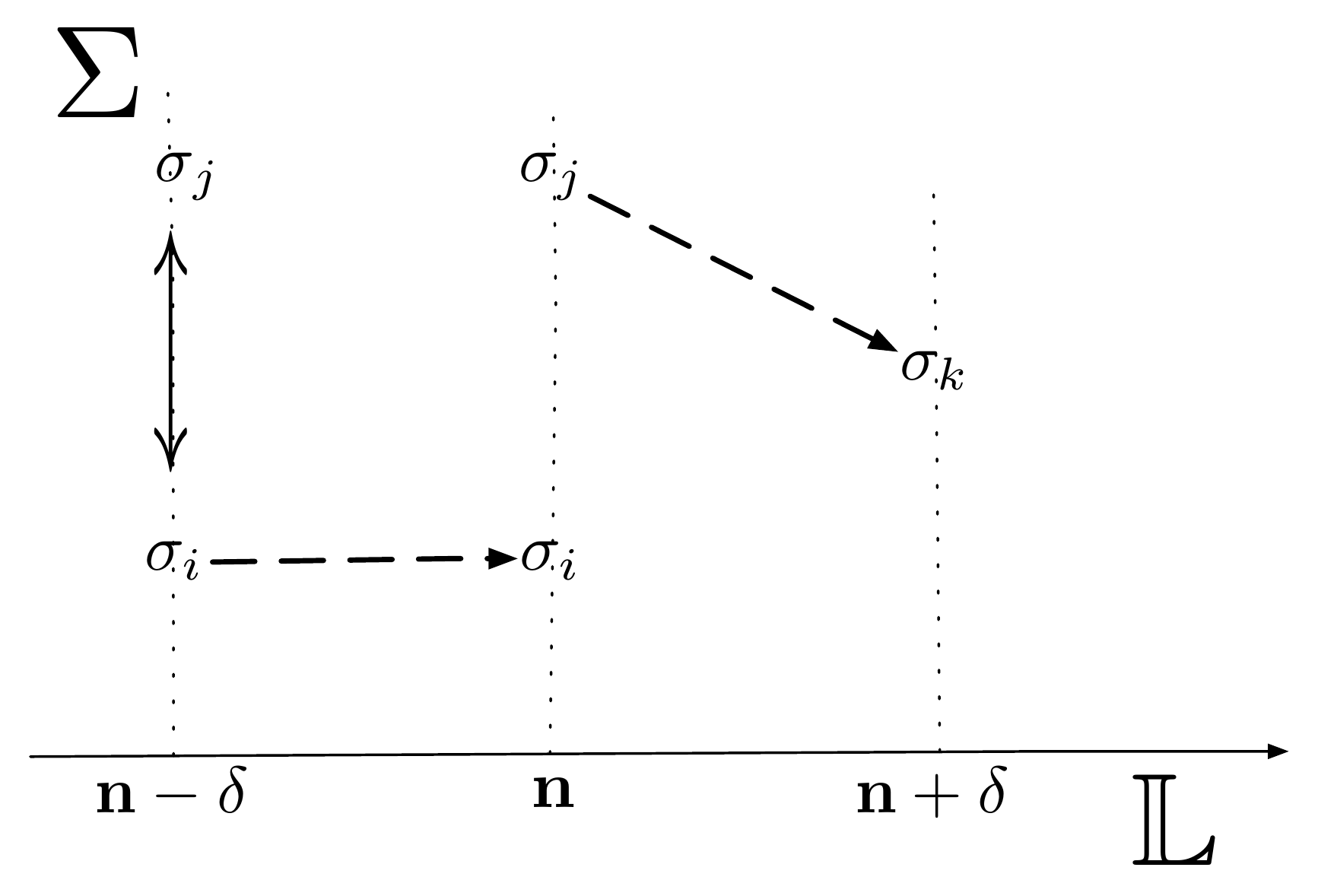} 
   \caption{A simple description of the the product $\LL\times\Sig$ and the graph links. Different types of state transitions result in vertical, horizontal or diagonal movements in the finite and infinite state sets.}
   \label{fig:Lattice-1D}
\end{figure}

\noi If one wants to include that the configuration space is the product $\LL\times\Sigma$, 
then the graph can be thought to be as in Figure \ref{fig:Lattice-1D}. 

\subsection{Formulation of the double limit $\tau\ra 0,\vde\ra 0$}
We are interested in the GME that results by taking the limits

\[\tau\ra 0\mbox{ and }\de_i\ra0\mbox{ for all $i$}.\]

\noi First we expand $P(\vn,\sig,t+\tau)$ up to the first order in $\tau$. This can be written as

\begin{equation}
\label{GME-step0}
\begin{array}{ll}
\displaystyle\frac{\pa P(\vn,\sig,t)}{\pa t}\tau=-P(\vn,\sig,t)+\sum_{\sig'\in\Sig}\int_0^t\der t'\,\Ld^*_{\sig\sig'}[\vn,\vde,\tau,t-t']\,P(\vn,\sig',t')+\\[4mm]
\displaystyle+\sum_{\sig'\in\Sig}\int_0^t\der t'\Kd^T_{\sig\sig'}[\vn,\vde,\tau,t-t']P(\vn,\sig',t')+o(\tau)
\end{array}
\end{equation}

or equivalently

\begin{equation}
\label{GME-step1}
\begin{array}{ll}
\displaystyle\frac{\pa P(\vn,\sig,t)}{\pa t}=\frac{1}{\tau}\, \int_0^t\der t'\,(\Ld^*_{\sig\sig}[\vn,\vde,\tau,t-t']-\de(t-t'))P(\vn,\sig,t') \\[4mm]
\displaystyle+\frac{1}{\tau}\,\sum_{\sig'\neq\sig}\int_0^t\der t'\,\Ld^*_{\sig\sig'}[\vn,\vde,\tau,t-t']\,P(\vn,\sig',t')
\\[4mm] 
\displaystyle +\frac{1}{\tau}\,\sum_{\sig'\in\Sig}\int_0^t\der t'\Kd^T_{\sig\sig'}[\vn,\vde,\tau,t-t']P(\vn,\sig',t')+o(\tau).
\end{array}
\end{equation}

\noi To proceed further it is necessary to study the following three limits:

\begin{equation}
\label{Term1}
\lim_{\tau\ra0,\vde\ra 0}\frac{1}{\tau}\int_0^t\der t'\,(\Ld^*_{\sig\sig}[\vn,\vde,\tau,t-t']-\de(t-t'))P(\vn,\sig,t'),
\end{equation}

\begin{equation}
\label{Term2}
\lim_{\tau\ra0,\vde\ra 0}\frac{1}{\tau}\sum_{\sig'\neq\sig}\int_0^t\der t'\,\Ld^*_{\sig\sig'}[\vn,\vde,\tau,t-t']\,P(\vn,\sig',t'),
\end{equation}

\begin{equation}
\label{Term3}
\lim_{\tau\ra0,\vde\ra 0}\frac{1}{\tau}\sum_{\sig'\in\Sig}\int_0^t\der t'\Kd^T_{\sig\sig'}[\vn,\vde,\tau,t-t']P(\vn,\sig',t').
\end{equation}

\subsection{Multiscale analysis: simplified assumptions}
The study of the limits (\ref{Term1}), (\ref{Term2}) and (\ref{Term3}) in this general form is very difficult.  In order to proceed and to analyse equation (\ref{discrete_GME}) some simplifying assumptions are in order. 
We shall consider two main sets of such assumptions which identify two classes of systems that  are
called  \emph{Infinite MC coupled with finite MC} and \emph{Infinite MC coupled with finite CTRW}, respectively. We  introduce them in this order and simultaneously discuss the continuum limit procedure.

\subsubsection*{Infinite MC coupled with finite MC} The first set of assumptions is:

\begin{itemize}
\item[(A1)] On $\LL$, we have $\de_i=\de$ for all $i$.
\item[(A2)] Each waiting time is exponentially distributed.
\item[(A3)] Each $\Ld^*_{\sig\sig'}$ is the adjoint of a generator of a Markov process valued in $\LL$.  
\item[(A4)] For fixed $\vn,\vde$, the transpose of the kernel $\Kd^T$ generates a Markov chain on $\Sig$.
\end{itemize}

\noi Under these conditions we have

\[\Ld^*_{\sig\sig'}[\vn,\vde,\tau,t-t']=\Ld^*_{\sig\sig'}[\vn,\de]\,\de(t-t')\mbox{ and }
\Kd^T_{\sig\sig'}[\vn,\vde,\tau,t-t']=\Kd^T_{\sig\sig'}[\vn,\de]\,\de(t-t').\]

\noi Here $\de(.)$ denotes the Dirac delta distribution. The limits (\ref{Term1}), (\ref{Term2}), (\ref{Term3}) reduce respectively to

\begin{equation}
\lim_{\tau\ra0,\de\ra 0}\frac{1}{\tau}(\Ld^*_{\sig\sig}[\vn,\de]-1)P(\vn,\sig,t),
\label{Term1A}
\end{equation}

\begin{equation}
\lim_{\tau\ra0,\de\ra 0}\frac{1}{\tau}\sum_{\sig'\neq\sig}\Ld^*_{\sig\sig'}[\vn,\de]P(\vn,\sig',t),
\label{Term2A}
\end{equation}

\begin{equation}
\lim_{\tau\ra0,\de\ra 0}\frac{1}{\tau}\sum_{\sig'\in\Sig}\Kd^T_{\sig\sig'}[\vn,\de]P(\vn,\sig',t).
\label{Term3A}
\end{equation}

\noi We can give a meaning to these limits by assuming that the two scales $\tau$ and $\de$ go to zero in a 
prescribed manner. A typical interesting regime is the diffusive one, namely when $\de^2/\tau\simeq D>0$, with $D$ being the diffusion coefficient.  Note that the limit process transforms the lattices $\LL$ into $\R^N$ into a limit state space given by 

\[\R^N\times\Sigma.\]

The continuum limit is based on the approximation method developed by Trotter in \cite{Trotter}, later also worked into \cite{Pazy}, \cite{Kurtz}. We shall 
now outline this approach. The ME is in general constructed as an operator acting on probability measures. One has to observe that the natural setting to construct the continuum 
limit is the space of functions, rather than the space of measures. We have seen that 
the ME is constructed by fixing the space-time scales $\vde$ and $\tau$. Let us introduce an 
index to enumerate the scales: $\vde_n$, $\tau_n$. The $n$th scale corresponds to the lattice $\LL_n$. We denote the result by $\X_n=\ell^\infty(\LL_n,\R^M)$, with norm

\[\|\phi\|_n=\sup_{\bk\in\Z^N}|\phi(\bk)|, \]

where $\phi(\bk)=\phi(k_1,...,k_N)$ is an element in the set $\X_n$. Each $\X_n$ is 
a Banach space and can be see as an "approximation" of $\X=C^0(\R^N,\R^M)$. In fact we can define the projection

\begin{equation}
\begin{array}{cc}
&\Pro_n:\X\mapsto\X_n,\\
&f\mapsto \Pro_n(f).
\label{projection}
\end{array}
\end{equation}

In particular, for any $f\in C^0(\R^N,\R^M)$, we have $P_n(f)(\bk)=f(\bk\vde_n)=f(k_1\de_n,...,k_N\de_n)$. The following  properties hold:

\begin{itemize}
\item[(i)] $\|\Pro_n\|_n\leq 1$,
\item[(ii)] $\lim_{n\ra\infty}\|\Pro_n(f)\|_n=\|f\|_0$ for all $f\in C^0(\R^N,\R^M)$.
\end{itemize} 

\noi See \cite{Pazy} for more details. Using \cite{Trotter}, we state a condition defining whether  a sequence of functions in the collection $\{\X_n\}_n$ of functions in $\X_n$ approximate a function in $\X$:

\bdefi
\label{WA}
Let $f_n\in\X_n$. The sequence $(f_n)$ converges to $f\in\X$ if and only if

\[\lim_{n\ra\infty}\|\Pro_n(f)-f_n\|_n=0.\]

This convergence is denoted here by $f_n\app f$.
\edefi

\noi This allows us to define the continuum limit:

\bdefi[Continuum limit of operators]
\label{ContinuumLimit}
Let $\Ld_n:\X_n\mapsto\X_n$. The sequence $(\Ld_n)$ of linear operators has a
\emph{continuum limit} $\Lop:\X\mapsto\X$ if an only if there exist a choice of $\vde_n$ 
and $\tau_n$ such that $\vde_n\ra 0$, $\tau_n\ra 0$ and 

\begin{equation}
\lim_{n\ra\infty}\|\Ld_n\Pro_n(f)-\Pro_n\Lop(f)\|_n=0\mbox{, for all $f\in\X$.}
\label{asymp}
\end{equation}

As before this limit is denoted by $\Ld_n\app\Lop$. The domain of $\Lop$ is formed by all $f\in\X$ such that the sequence $\Ld_n\Pro_n(f)\in \X_n$ 
converges.
 \edefi


\brem
It is worth emphasising that the continuum limit of an operator is not unique. In fact relation between $\vde_n$ and $\tau_n$ is  crucial in definition \ref{ContinuumLimit}. We shall see in the examples that the scaling relations among the parameters in $\Ld_n$ identify the possible continuum limits. 
\erem

\noi For every fixed $n$ the dual of $\X_n$ is the space $\X_n^*$, formed by the measures $P$ such that

\begin{equation}
\langle P,\phi\rangle_n=\sum_{\bk}\phi(\bk)( \cdot ) P(\bk)=\sum_{\bk}\left(\sum_{i=1}^M\phi_i(\bk)P_i(\bk)\right)
\label{duality}
\end{equation}

is finite. The ME is defined on the dual space $\X^*_n$. Using the pairing (\ref{duality}) one can transfer the ME to be defined on $\X_n$ by using

\[\langle \Ld_n^*P,\phi\rangle_n=\langle P,\Ld_n \phi\rangle_n\mbox{ and }\langle \Kd_n^TP,\phi\rangle_n=\langle P_n,\Kd \phi\rangle_n.\]

On the Banach space $\X$ the standard duality is given by

\begin{equation}
\langle \rho,f\rangle=\int_{\R^N}\der\bx f(\bx)( \cdot ) \rho(\bx)=\int_{\R^N}\der\bx\sum_{i=1}^Mf_i(\bx)\rho_i(\bx).
\label{Xduality}
\end{equation}

Given the continuum limits $\Lop$ and $\Kop$ we can therefore define their adjoints

\[\langle \rho,\Lop \,f\rangle=\langle \Lop^*\,\rho,f\rangle\mbox{ and }\langle \rho,\Kd f\rangle=\langle \Kd^T\rho,f\rangle.\]

We now give some basic examples. First let us state

\bdefi
Let $\LL=\de\,\N$, and let $\phi$ be $\phi:\LL\ra\R$. Then the couple of operators $\Dop^\pm$ acts on $\X_n$:

\[\Dop^\pm(\phi(k))=(\Eop^\pm-\id)(\phi(k))=\phi(k\pm \de_n)-\phi(k).\]
\edefi

One can easily show that $(\Dop^+)^*=-\Dop^-$, namely

\[\langle \Dop^+P,\phi\rangle_n=-\langle P,\Dop^-\phi\rangle_n\mbox{, for every $n$.}\]

Also clearly the operator $\Dop^++\Dop^-$ is symmetric. We compute the continuum limit of $\Ld_n^+=(1/\tau_n)\Dop^+$ and take a sequence $P_nf$ with $f\in \X$. The aim is to compute 
 
\[\Ld^+\Pro_n(f)(k)=\frac{1}{\tau_n}\,(f(k\de_n+\de_n)-f(k\de_n)).\]

 This can be rewritten as
 
\[\Ld^+\Pro_n(f)(k)=\Pro_n\left(\frac{1}{\tau_n}\,(f(x+\de_n)-f(x))\right).\]

For $n$ large enough, $\de_n,\tau_n$ are arbitrary small. Taking $f$ in a suitable dense subspace of $\X$ we can write

\[\Ld^+\Pro_n(f)(k)=\Pro_n\left(\frac{\de_n}{\tau_n}\,\frac{\pa f}{\pa x}+\frac{\de^2_n}{\tau_n}\,\frac{\pa^2 f}{\pa x^2}\right)
+\frac{1}{\tau_n}\,o(\de_n^2).\]

Now (\ref{asymp}) can be verified by taking the limit. We have that

\[\Lop^+=c\,\frac{\pa}{\pa x}\mbox{ for $\de_n/\tau_n\ra c>0$.}\]

Once the continuum limit of $\Ld$ is constructed on $\X$ we can take the adjoint operator 
with respect (\ref{Xduality}).  Then, for example, $(\Lop^+)^*=(c\pa/\pa x)^*=-c(\pa/\pa x)$. In this sense we assume  that the limits (\ref{Term1A}), (\ref{Term2A}) and (\ref{Term3A})  are

\begin{equation}
\frac{1}{\tau}(\Ld^*_{\sig\sig}[\vn,\de]-1)P(\vn,\sig,t)\app\Lop^*_{\sig\sig}[\vx]\rho_\sig(\vx,t),
\label{limit1A}
\end{equation}

\begin{equation}
\frac{1}{\tau}\sum_{\sig'\neq\sig}\Ld^*_{\sig\sig'}[\vn,\de]P(\vn,\sig',t)
\app\sum_{\sig\neq\sig'}\Lop^*_{\sig\sig'}[\vx]\rho_{\sig'}(\vx,t),
\label{limit2A}
\end{equation}

\begin{equation}
\frac{1}{\tau}\sum_{\sig'\in\Sig}\Kd^T_{\sig\sig'}[\vn,\de]P(\vn,\sig',t)\app\sum_{\sig'\in\Sig}\Kop^T_{\sig\sig}[\vx]\rho_{\sig'}(\vx,t),
\label{limit3A}
\end{equation}

where $\Lop^*$ is a matrix with entries differential operators and $\Kop^T$ is the transpose of the infinitesimal generator of a finite Markov chain on $\Sigma$. An interesting further simplification is obtained if processes of type (P3) do not occur. This implies that 

\begin{itemize}
\item[(i)] $\Ld^*_{\sig\sig'}\equiv0$ for $\sig\neq\sig'$,
\item[(ii)] $\Ld^*_{\sig\sig}$ are Fokker-Planck operators.
\end{itemize} 

In this case we have that the degrees of freedom are represented by $\vx$ diffuse in $\R^N$, while 
 the discrete states $\sig$'s evolve in $\Sig$ according to a finite Markov chain generated by $\Kd$.\\[4mm]
 
 We  now look at a second set of  assumptions:

\subsubsection*{Infinite MC coupled with CTRW} The second set of assumptions is

\begin{itemize}
\item[(A1)] On $\LL$, $\de_i=\de$ for all $i$.
\item[(B2)] The waiting times $\psi_{\sig, n_i}(t;\tau,\de_i)$ are exponentially distributed, independent of $\vde$ and $\tau$.
\item[(B3)] Each $\Ld_{\sig\sig'}$ is a generator of a Markov process valued in $\LL$.  
\item[(B4)] For fixed $\vn,\vde$ the kernel $\Kd$ generates a continuous-time random walk (CTRW) on $\Sig$.
\end{itemize}

\noi Under these conditions we have

\[\Ld_{\sig\sig'}[\vn,\vde,\tau,t-t']=\Ld_{\sig\sig'}[\vn,\de]\,\de(t-t')\mbox{ and }
\Kd_{\sig\sig'}[\vn,\vde,\tau,t-t']=\tau\,\Kd_{\sig\sig'}[\vn,\de,t-t'].\]

\noi The possible form of the limit $\tau\ra 0,\de\ra 0$ can be written as

\begin{equation}
\frac{1}{\tau}(\Ld^*_{\sig\sig}[\vn,\de]-1)P(\vn,\sig,t)\app\Lop^T_{\sig\sig}[\vx]\rho_\sig(\vx,t),
\label{Term1B}
\end{equation}

\begin{equation}
\frac{1}{\tau}\sum_{\sig'\neq\sig}\Ld^*_{\sig\sig'}[\vn,\de]P(\vn,\sig',t)
\app\sum_{\sig\neq\sig'}\Lop^*_{\sig\sig'}[\vx]\rho_{\sig'}(\vx,t),
\label{Term2B}
\end{equation}

\begin{equation}
\frac{1}{\tau}\sum_{\sig'\in\Sig}\Kd^T_{\sig\sig'}[\vn,\de]P(\vn,\sig',t)
\app\sum_{\sig'\in\Sig}\int_0^t\der t'\,\Kop^T_{\sig\sig}[\vx,t-t']\rho_{\sig'}(\vx,t-t').
\label{Term3B}
\end{equation}


\brem
The main reason to use Trotter \emph{approximation} is that it was proven in \cite{Trotter}, \cite{Kurtz} and \cite{Pazy} that if each operator $\Ld_n$ defined on $\X_n$ is an infinitesimal  generator of a (strongly continuous contraction) semigroup $T_n(t)$, then the limit operator 
$\Lop$ is also the  generator of (strongly continuous contraction) semigroup $T(t)$ on $\X$. This fact guarantees us that  the continuum limit procedure produces a meaningful approximation of the real dynamics. For more details see \cite{LM1} and \cite{LM2}.
\erem

\section{General Fokker-Planck equation and the adiabatic condition}
Upon the condition that limits (\ref{Term1A}), (\ref{Term2A}) and (\ref{Term3A}) exist, the probability density  $\rho(\vx,t)$ satisfies a general Fokker-Planck equation of the form:
 
 \begin{equation}   
\frac{\pa\rho(\vx,t)}{\pa t}=\Lop^*[\vx]\circ\rho(\vx,t)+\Kop^T[\vx]\,\rho(\vx,t).
\label{gen-FP1}
\end{equation}

\noi If the limits (\ref{Term1B}), (\ref{Term2B}) and (\ref{Term3B}) exist,  equation (\ref{gen-FP1}) 
can be modified into

\begin{equation}   
\frac{\pa\rho(\vx,t)}{\pa t}=\Lop^*[\vx]\circ\rho(\vx,t)+\Kop^T[\vx,t]*\rho(\vx,t),
\label{gen-FP2}
\end{equation}

\noi where $\Kop[\vx,t]*\rho(\vx,t)$ is a time convolution.

\subsubsection*{Adiabatic condition}
The construction of the continuum limit involves a choice in which way $\de_n$ and $\tau_n$ 
tend to zero. This implies that the operators $\Lop$ and $\Kop$ may have a pre-factor which is a function of $\de_n$ and $\tau_n$. These coefficients determine the different time scales at which  the operators $\Lop$ and $\Kop$ influence the dynamics. We shall see in the examples that the continuum limit procedure often results in an FPE of the form

\begin{equation}   
\frac{\pa\rho(\vx,t)}{\pa t}=\Lop^*[\vx]\circ\rho(\vx,t)+\frac{1}{\eps}\,\Kop^T[\vx]\,\rho(\vx,t),
\label{gen-FP1-ad}
\end{equation}

where $\eps=\eps(\vde_n,\tau_n)\in (0,1]$.  This corresponds to the assumption that the Markov chain dynamics is faster than the diffusion process.\\

The condition  $\eps=\eps(\vde_n,\tau_n)$ small is called \emph{adiabatic},  because it  determines 
a separation between the dynamics of $\Kd$ and of $\Ld$. In fact for $\eps=0$  the dynamics of the system is dominated by the Markov chain at equilibrium. This is given by a linear combination of stationary measures of the Markov chain defined by

\[\mu(\vx)\,\Kd[\vx]=0.\] 

Intuitively one can see that for $\eps$ small the time evolution of the whole system will organise itself around the steady state of the Markov chain. In order to introduce the result we need to define:

\bdefi
Let $C_\Kop\subset\R^M$ be the convex cone of stationary measures of $\Kop[\vx]$. 
\edefi

\noi Consider the operator

\begin{equation}
\langle\be_\mu,\Lop^*\vmu(\vx)( \cdot )\rangle=\sum_{\sig',\sig\in\Sig_\mu}\Lop_{\sig'\sig}[\vx]\mu_\sig(\vx)\,( \cdot )~~~\mbox{where $\Sig_\mu=\{\sig\in\Sig:\mu_\sig(\vx)\neq 0\}$.}
\label{basic-cond}
\end{equation}

\noi In \cite{LM1} the following has been proved: 

\bth
Upon the condition that 

\begin{equation}
\frac{\pa f(\vx,t)}{\pa t}=\langle\be_\mu,\Lop^*\vmu(\vx)f(\vx,t)\rangle + F(\vx,t)
\end{equation}

 yields a probability density  which is differentiable w.r.t. $\bx$ and $t\in [0,T_0]\subset [0,T]$,
for any smooth initial data and smooth $F(\vx,t)$,  equation  (\ref{gen-FP1-ad}) can be solved by an asymptotic expansion of the form

\[\rho_\eps(\vx,t)=\sum_{k=0}\eps^n\,\rho^{(n)}(\vx,t).\]
\eth

\bpf Here we only present a sketch of the proof.  This is essentially based on the adiabatic theory 
developed in \cite{Pavliotis}. In \cite{LM1} we  show that a solution of the ????? can be constructed asymptotically in $\eps$. The main steps of the proof are the following:

\begin{enumerate}
\item Take $\vmu(\vx)\in C_\Kop$ and fix an initial $\rho(\vx,0)$ such that $\supp{(\rho(\vx,0))}\subset\supp{(\vmu(\vx))}$.
\item Consider an expansion of the form: $\rho_\eps(\bx,t)=\sum_{n=0}^{m^*}\eps^n\,\rho^{(n)}(\bx,t)$.
\item Construct the equation at each order $n$.
\item Decompose each $\rho^{(n)}(\bx,t)$ using the projection $\bPi_\vmu$:

\[ \rho^{(n)}(\bx,t)=\vxi^{(n)}(\vx,t)+\vmu(\bn)\,f^{(n)}(\vx,t),\]

where

\[\vxi^{(n)}(\bx,t)=\bPi_\vmu(\rho^{(n)}(\vx,t)),~~~f^{(n)}(\vx,t)
=\langle\be_\mu,\rho^{(n)}(\vx,t)\rangle=\sum_{\sig\in\Sig_\mu}\rho_\sig^{(n)}(\vx,t).\]

\item Construct the hierarchy of equations:
for $n=0$:

\begin{equation}
\left\{\begin{array}{ll}
\displaystyle\vxi^{(0)}(\vx,t)=0\\[4mm]
\displaystyle\frac{\pa f^{(0)}(\vx,t)}{\pa t}=\langle\be_\vmu,\Lop^*[\vx](\vmu(\vx)\,f^{(0)}(\vx,t))\rangle
\end{array}
\right. ,
\label{dual-step0}
\end{equation}

and for $n\geq 1$ we have

\begin{equation}
\left\{\begin{array}{ll}
\displaystyle\vxi^{(n)}(\vx,t)=(\Kop^T_\vmu)^D\bI_\vmu\left[\frac{\pa \vxi^{(n-1)}(\vx,t)}{\pa t}-\Lop^*(\vxi^{(n-1)}(\vx,t)+\vmu(\vx)\,f^{(n-1)}(\bn,t))\right]\\[4mm]
\displaystyle\frac{\pa f^{(n)}(\vx,t)}{\pa t}=\langle\be_\vmu,\Lop^*(\vmu(\vx)\,f^{(n)}(\vx,t))\rangle+\langle\be_\vmu,\Ld^*(\vxi^{(n)}(\vx,t))\rangle,
\end{array}
\right. ,
\label{dual-stepn}
\end{equation}

where $(\Kop^T_\vmu)^D$ is the Drazin inverse of $\Kop$.
\item The evaluation of the remainder of the asymptotic series is then carried out as
in \cite{LM1}.
\end{enumerate}
\epf

\noi If the adiabatic condition holds then the asymptotic approximation can also be constructed for a ME of the form

\begin{equation}   
\frac{\pa P(\vn,t)}{\pa t}=\Ld^*[\vn]\circ P(\vn,t)+\frac{1}{\eps}\,\Kd^T[\vn]\,P(\vn,t).
\label{gen-ME1-ad}
\end{equation}
\noi In this case the hierarchy of equations has to be modified accordingly. For the benefit of the reader 
we include the the hierarchy of equations.\\
For $n=0$:

\begin{equation}
\left\{\begin{array}{ll}
\displaystyle\xi^{(0)}(\vn,t)=0\\[4mm]
\displaystyle\frac{\pa f^{(0)}(\vn,t)}{\pa t}=\langle\be_\mu,\Ld^*[\vn](\mu(\vn)\,f^{(0)}(\vn,t))\rangle
\end{array}
\right.
\label{dual-step0N}
\end{equation}

and for $n\geq 1$ we have

\begin{equation}
\left\{\begin{array}{ll}
\displaystyle\xi^{(n)}(\vn,t)=(\Kd^T[\vn]_\mu)^D\bI_\mu\left[\frac{\pa \xi^{(n-1)}(\vn,t)}{\pa t}-\Ld^*[\vn](\xi^{(n-1)}(\vn,t)+\mu(\vn)\,f^{(n-1)}(\vn,t))\right]\\[4mm]
\displaystyle\frac{\pa f^{(n)}(\vn,t)}{\pa t}=\langle\be_\mu,\Ld^*[\vn](\mu(\bn)\,f^{(n)}(\vn,t))\rangle+\langle\be_\mu,\Ld[\vn]^*(\xi^{(n)}(\vn,t))\rangle.
\end{array}
\right.
\label{dual-stepnN}
\end{equation}

\subsection{Reduction of the Interaction Graph}
Equations (\ref{gen-FP1}), (\ref{gen-FP2}) allow again an interpretation  in terms of graphs. In fact the operators $\Lop^*$ and $\Kop^T$  describe the rates at which the transitions of type 

 \[(\vx,\sig)\ra(\vx',\sig')\]
 
 occur.  We can define:
 
 \bdefi[Interaction Graph for the FPE]
 We term $\vI_G(V,E)$ the graph whose vertex set is $V=\R^N\times\Sig$ and edge set is
$E(\Lop,\Kop)$ where the directed link $\ve_{(\vx,\sig),(\vx',\sig')}$ is present if 
$\Lop$ and $\Kop$ allow the transition $(\vx',\sig')\mapsto (\vx',\sig')$. 
 \edefi
 
 \brem
 One can think that 
 a representation of this graph can be obtained by looking at the Figures \ref{fig:GG} and 
 and \ref{fig:Lattice-1D},  where instead of $\vn$'s there are $\vx$'s. This is only partly true because 
it might happen that the continuum limit procedure  removes some states and is thus changing the graph topology.
 \erem
 
\noi We can observe different levels of reduction and simplification of the Interaction Graph $\I_G$. If $\Lop^*$ is such that

\begin{itemize}
\item[(i)] $\Lop^*_{\sig\sig'}\equiv0$ for $\sig\neq\sig'$,
\item[(ii)] $\Lop^*_{\sig\sig}$ are Fokker-Panck operators,
\end{itemize} 

then the only possible processes have the form:

 \begin{itemize}
 \item[(a)] $(\vx,\sig)\ra(\vx',\sig)$,
 \item[(b)] $(\vx,\sig)\ra(\vx,\sig')$.
 \end{itemize}
 
\begin{figure}[htbp] 
   \centering
   \includegraphics[scale=0.4]{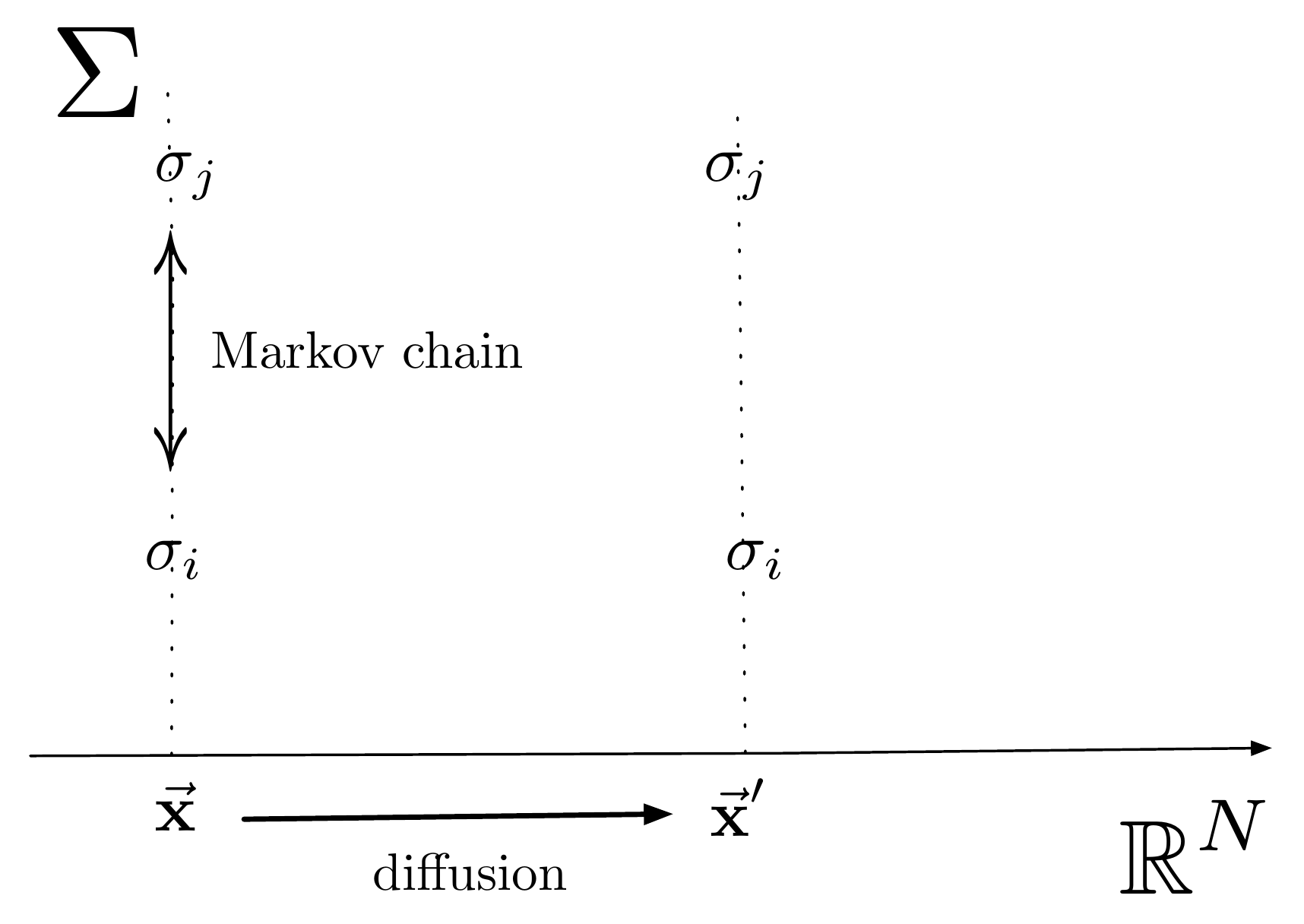} 
   \caption{A Markov chain "fibered" over $\R^N$.}
   \label{fig:Diff_MC-1D}
\end{figure}

In particular  (a) corresponds to a \emph{diffusive} Markov process and (b) corresponds to a finite Markov chain. We can think to the following scheme: on each point of $\R^N$ where the diffusion take place there is a  "fibered" Markov chain whose transition rates are functions of $\vx$, see  Figure \ref{fig:Diff_MC-1D}.

This reduction takes place also in equation (\ref{gen-ME1-ad}).\\[2mm]

\subsubsection{The average dynamics and its Interaction Graph}
Let us now consider the zero order approximation of the $\eps$ expansion. 
This is given by 

\begin{equation}
\frac{\pa f^{(0)}(\vx,t)}{\pa t}=\langle\be_\vmu,\Lop^*[\vx](\vmu(\vx)\,f^{(0)}(\vx,t))\rangle.
\label{average-dynamics}
\end{equation}

This is called \emph{average dynamics}. The average dynamics is a Liouville  equation for a \emph{deterministic vector field} given by

\begin{equation}
\frac{\der\vx(t))}{\der t}=-\sum_{\sig'\in\Sig_\mu}\sum_{\sig\in\Sig_\mu}\Ld_{\sig'\sig}^*[\vx(t)]\mu_\sig(\vx(t)).
\label{average-vf}
\end{equation}

\noi We can give a description of the average vector field by using the notion of an interaction graph. We define:

\bdefi[Interaction Graph for deterministic dynamics]
For a given vector field $\vcX(\vx)$, the \underline{Interaction Graph} $\vI_G^\sig\doteq\vI_G(\vcX)$  is 
 the couple $(V,E_\vcX)$ where:

\begin{itemize}

\item[(i)] $V$ is the set equal to the collection $\{x_1,...,x_N\}$,


\item[(ii)] $E_\vcX$ is the set of edges $\ve_{ij}$. The edge $\ve_{ij}$ is associated to the couple of vertices $(x_i,x_j)$   if 

\[\frac{\pa \cX_i(\vx)}{\pa x_j}\mbox{ is not identically zero.}\] 

\item[(iii)] The edge $\ve_{ij}$ is directed from $j$ to $i$.
\end{itemize}
\edefi

\noi Note that for each fixed $\sigma\in \Sig$ we can associate a vector field 

\begin{equation}
\cX^{(\sig)}(\vx(t))\doteq-\sum_{\sig'\in\Sig_\mu}\Ld_{\sig'\sig}^*[\vx(t)],
\label{Xsigma}
\end{equation}

\begin{figure}[htbp] 
   \centering
   \includegraphics[scale=0.4]{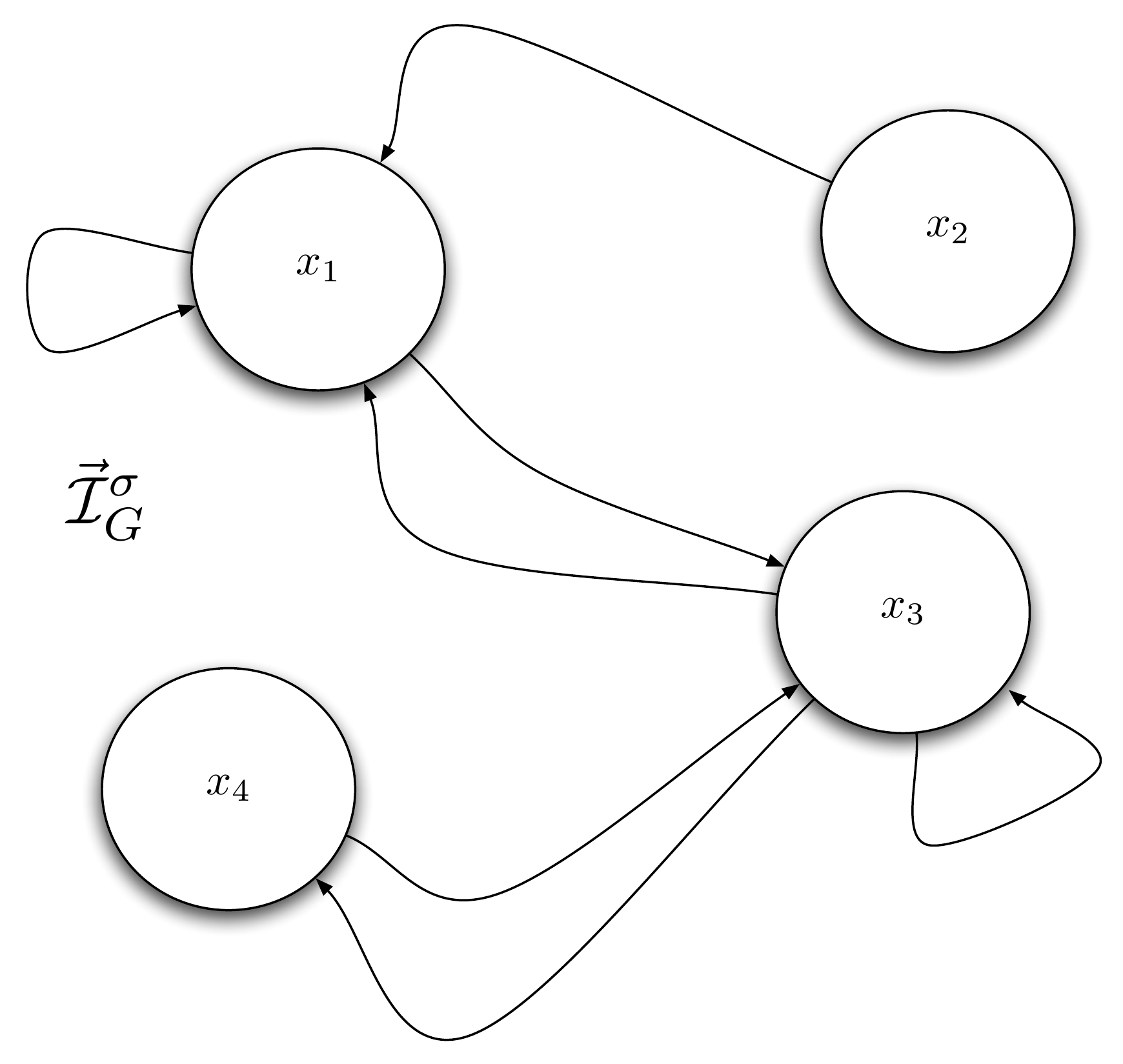} 
   \caption{The Interaction Graph for a fixed  state $\sig\in\Sig_\mu$ in the Markov chain}
   \label{fig:IGsigma}
\end{figure}

\noi and therefore an interaction graph $\vI_G({\vcX^{(\sig)}})$. A graphical description is presented in Figure \ref{fig:IGsigma}.

\noi It is simple to note that  the average vector-field (\ref{average-vf}) can also be written as

\begin{equation}
\frac{\der\vx(t))}{\der t}=\vcX_\vmu(\vx)\doteq-\sum_{\sig\in\Sig}\cX^{(\sig)}(\vx)\mu_\sig(\vx).
\label{average-vf2}
\end{equation}

This vector field is the average of  all $\vcX^{(\sig)}$ taken against the invariant measure $\vmu(\vx)$. This implies that  the associated interaction graph $\I_G(\vcX_\mu)$ has a new structure. The vertices $V=\{x_1,...,x_N\}$ will not contain reference to the specific Markov chain state $\sig$ and new edges will appear as a result of new interaction terms resulting from the averaging procedure.

\begin{figure}[htbp] 
   \centering
   \includegraphics[scale=0.4]{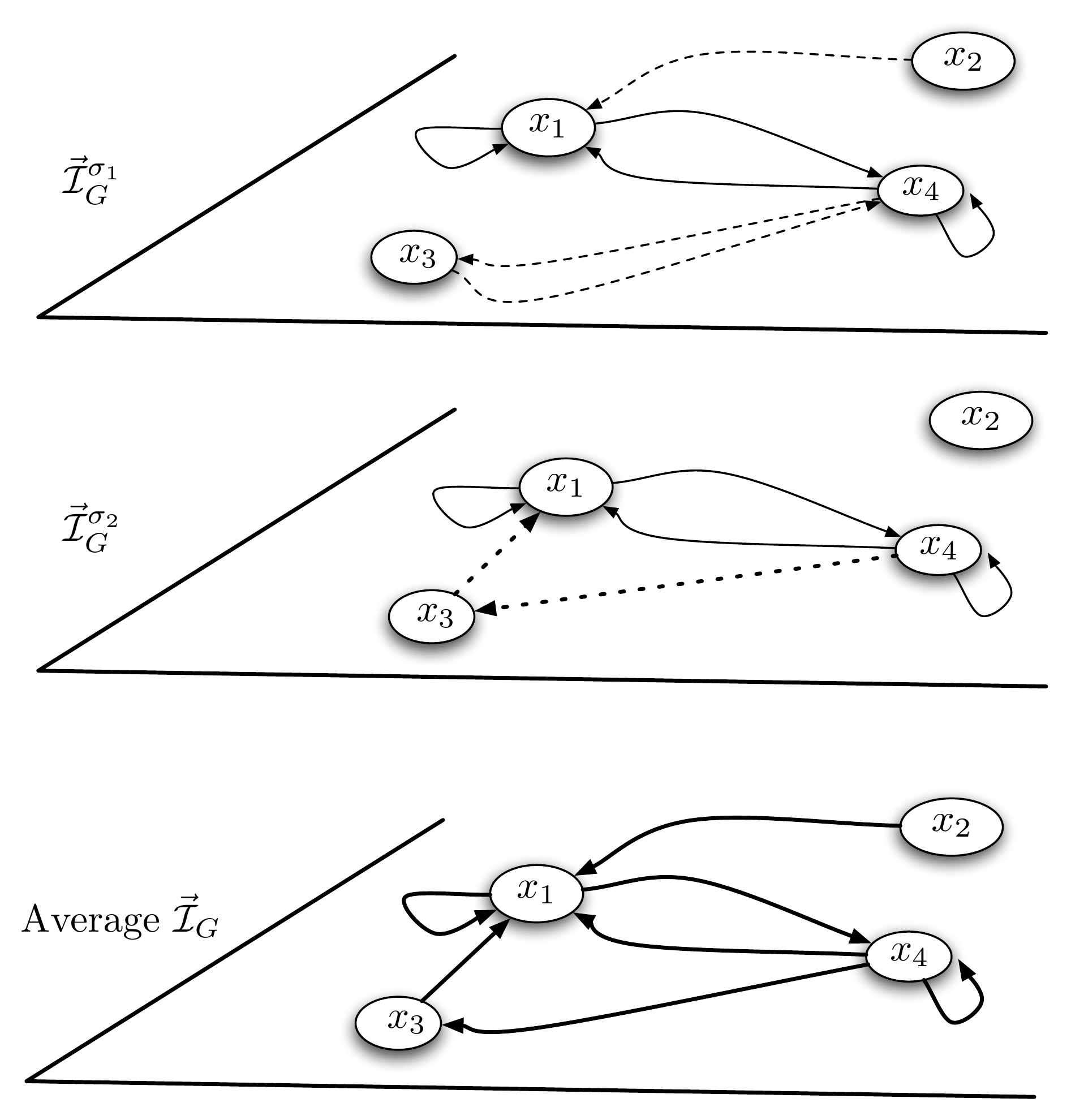} 
   \caption{An example where $\Sig_\mu=\{\sig_1,\sig_2\}$ and $\vx=(x_1,x_2,x_3,x_4)$. The averaging process gives rise to an new Interaction Graph depending on the combination of $\vI_G^{\sig_1}$ and $\vI_G^{\sig_2}$. The edges that are present in 
   $\vI_G^{\sig_1}$ (dashed arrow lines) are eventually also present in $\vI_G$, and so do 
   the edges (point arrow lines) present in $\vI_G^{\sig_2}$. Note that there might occur compensations that delete some links.}
   \label{fig:IGaverage}
\end{figure}

\brem
It is worth noting that if $\{\vcX^{(\sig)}\}_{\sig\in\Sig}$ are all  polynomial vector fields with integer coefficients, then the average vector field can be rewritten as 

\begin{equation}
\vcX(\vx)=N\,\vec{\nu}(\vx).
\label{Nnu}
\end{equation}

Here $\vec{\nu}(\vx)$ is a vector field whose entries are in general rational functions.  At this 
point the analysis could proceed along the lines of exploiting all the graphic structure present in (\ref{Nnu}). A review of the results in this direction is presented in \cite{Mirela}.  
\erem

\noi In the next section we shall look at some simple examples. We restrict ourselves to the case of models where all the waiting times are exponentially distributed. Therefore we consider an infinite Markov coupled to a finite one and we show that also in this simple setting many interesting properties and question arise.

\section{Examples}
Let us first fix our setting. To avoid cumbersome notation we drop the index from $\de_n$ and $\tau_n$ and 
the projection $\Pro_n$. We shall consider the following cases

\begin{itemize}
\item[(i)] $\LL=\de\,\Z$ a random walk with an internal two-state space $\Sig$ (switch).
\item[(ii)] $\LL=\de\,\N$ a single particle regulating a two-state  switch.  
\item[(iii)] $\LL=\de\,\N\times\de\,\N$ two particles $A$ and $M$. $A$ regulates a two-state switch which in turn regulates $M$.  
\end{itemize}

The finite set $\Sig$ is always a collection of states modelling "molecular" operators and for this reason the elements of  $\Sig$ will be denoted by $O_\sig$ with $\sig=1,...,M$. The processes will be given in terms of reactions that will be interpreted as reaction rates. In the examples we formulate the problem using the reactions to construct a Master equation of the following form

\begin{equation}
\frac{\pa P_\sig(\vn,t)}{\pa t}=\frac{1}{\tau}\,\sum_{\sig'\in\Sig}\Ld_{\sig\sig'}[\vn,\de]\,P_{\sig'}(\vn,t)+\frac{1}{\tau}\,\sum_{\sig'\in\Sig}\Kd_{\sig\sig'}[\vn,\de]\,P_{\sig'}(\vn,t),
\label{approx-ME}
\end{equation}

where $\Ld_{\sig\sig'}[\vn,\de]$ is function of $\Eop_{n_i}^\pm$. In the various examples we want to illustrate how to define the continuum limit.  For equation (\ref{approx-ME})  the limit can be obtained by defining (\ref{limit1A}), (\ref{limit2A}) and (\ref{limit3A}). In particular we shall consider cases where

\begin{equation}
\frac{1}{\tau}\sum_{\sig'\in\Sig}\Kd_{\sig\sig'}[\vn,\de]P(\vn,\sig',t)\app\frac{1}{\eps}\sum_{\sig'\in\Sig}\Kop^*_{\sig\sig}[\vx]\rho_{\sig'}(\vx,t),
\label{Kcond}
\end{equation}

with $\eps=\eps(\de,\tau)$  for $\de,\tau\ra 0$. Upon this condition we shall show that a master equation has limit of the form (\ref{gen-FP1-ad}).

\subsection{Effective diffusion}
In the first example we consider a particle performing a random walk on $\LL=\de\,\Z$ with rates 
depending on an internal state $O_\sig\in\Sig=\{O_0,O_1\}$. The internal state dynamics is a 
Markov chain whose rates are dependent on the point where the particle is at time $t$. We assume exponentially distributed waiting times. The aim is to
show the various scaling regimes when $\de,\tau\ra 0$. In the adiabatic regime the motion of the particle will be given by an effective diffusion equation. The processes can be described through the following reactions

\[\begin{array}{lll}
\mbox{Diffusion with random rates}\\[2mm]
N \rlha[u^i(\de,\tau)]{v^i(\de,\tau)} N+1\mbox{ for $i=0,1$}\\[3mm]
O_0 \rlha[k^1(N,\de,\tau)]{k^0(N,\de,\tau)} O_1
\end{array}\]

The state a time $t$ is determined by the probability distribution

\[P(n,t)=(P_0(n,t),P_1(n,t)).\]

The Master equation is given by:

\begin{equation}
\left\{\begin{array}{llll}
\displaystyle\frac{\pa P_0(n,t)}{\pa t}=\overbrace{(u_0(\de,\tau)/\tau)\,(P_0(n-1)-P_0(n,t))+(v_0(\de,\tau)/\tau)(P_0(n+1,t)-P_0(n,t))}^{\mbox{particle motion}}+\\[3mm]
\displaystyle\overbrace{-(k^1(n,\de,\tau)/\tau)P_0(n,t)+(k^0(n,\de,\tau)/\tau)P_1(n,t)}^{\mbox{Markov chain: rates dynamics}}\\[4mm]
\displaystyle\frac{\pa P_1(n,t)}{\pa t}=\overbrace{(u_1(\de,\tau)/\tau)\,(P_1(n-1)-P_1(n,t))+(v_1(\de,\tau)/\tau)(P_1(n+1,t)-P_1(n,t))}^{\mbox{particle motion}}+\\[3mm]
\displaystyle\overbrace{+(k^1(n,\de,\tau)/\tau)P_0(n,t)-(k^0(n,\de,\tau)/\tau)P_1(n,t)}^{\mbox{Markov chain: rates dynamics}}
\end{array}\right.
\end{equation}

This can be rewritten as

\begin{equation}
\left\{\begin{array}{llll}
\displaystyle\frac{\pa P_0(n,t)}{\pa t}=(u_0(\de,\tau)/\tau)\Dop_n^-(P_0(n,t))+(v_0(\de,\tau)/\tau)\Dop_n^+(P_0(n,t))+\\[3mm]
\displaystyle-(k^1(n,\de,\tau)/\tau)P_0(n,t)+(k^0(n,\de,\tau)/\tau)P_1(n,t)\\[4mm]
\displaystyle\frac{\pa P_1(n,t)}{\pa t}=(u_1(\de,\tau)/\tau)\Dop_n^-(P_1(n,t))+(v_1(\de,\tau)/\tau)\Dop_n^+(P_1(n,t))+\\[3mm]
\displaystyle+(k^1(n,\de,\tau)/\tau)P_0(n,t)-(k^0(n,\de,\tau)/\tau)P_1(n,t)
\end{array}\right.
\end{equation} 
 
 \noi By applying (\ref{asymp}) $P_i(n,t)=\Pro(\rho_i(x,t))$, for $\de,\tau\ra 0$., we get
 
 \[
\begin{array}{ll}
(k^1(n,\de,\tau)/\tau)\Pro(\rho_0(x,t))=\Pro((k^1(x,\de,\tau)/\tau)\,\rho_{1}(x,t)),\\[4mm]
(k^0(n,\de,\tau)/\tau)\Pro(\rho_1(x,t))= \Pro((k^0(x,\de,\tau)/\tau)\,\rho_{0}(x,t))\\[4mm]
\end{array}\]

and

\[\begin{array}{ll}
\displaystyle (u_i(\de,\tau)/\tau)\Dop_n^-\Pro(\rho_{i}(x,t))\app (u_i(\de,\tau)/\tau)\,\Pro\left(\left[-\de\frac{\pa\rho_i}{\pa x}+\frac{\de^2}{2}\frac{\pa^2\rho_i}{\pa x^2}\right]\right)\\[4mm]
\displaystyle (v_i(\de,\tau)/\tau)\Dop_n^+\Pro(\rho_{i}(x,t))\app (v_i(\de,\tau)/\tau)\,\Pro\left(\left[\de\frac{\pa\rho_i}{\pa x}+\frac{\de^2}{2}\frac{\pa^2\rho_i}{\pa x^2}\right]\right)\\[4mm]
\end{array}
\]

\noi We now chose the scaling for $\de\ra0,\tau\ra 0$:

\[(u_i(\de,\tau)/\tau)\simeq u_i/\eps,~~(v_i(\de,\tau)/\tau)\simeq v_i/\eps,~~(k^0(x,\de,\tau)/(\de\tau))\simeq k^0(x)/\eps,~~(k^1(x,\de,\tau)/\tau)\simeq k^1(x)/\eps \]

for $i=0,1$ and some $\eps=\eps(\de,\tau)$. Then we can take the continuum limit and obtain: 

\begin{equation}
\left\{\begin{array}{llll}
\displaystyle\frac{\pa \rho_0(x,t)}{\pa t}=\frac{\de\,(v_0-u_0)}{\eps}\frac{\pa\rho_0(n,t)}{\pa x}+\frac{\de^2\,(v_0+u_0)}{\eps}\frac{\pa^2\rho_0(n,t)}{\pa x^2}+\\[3mm]
\displaystyle-\frac{k_1(x)}{\eps}\rho_0(n,t)+\frac{k^0(x)}{\eps}\rho_1(x,t)+\frac{1}{\tau}\,o(\de^2)\\[4mm]
\displaystyle\frac{\pa \rho_1(x,t)}{\pa t}=\frac{\de\,(v_1-u_1)}{\eps}\frac{\pa\rho_0(n,t)}{\pa x}+\frac{\de^2\,(v_1+u_1)}{\eps}\frac{\pa^2\rho_1(n,t)}{\pa x^2}+\\[3mm]
\displaystyle+\frac{k_1(x)}{\eps}\rho_0(n,t)-\frac{k^0(x)}{\eps}\rho_1(x,t)+\frac{1}{\tau}\,o(\de^2).\\[4mm]
\end{array}\right.
\end{equation} 

It is useful to check  the compatibility. First note that to simplify $o(\de^2)/\tau$ term we need that $\tau\ra 0$ such that 

\[\frac{1}{\tau}\,o(\de^2)\ra 0.\]

Next we  explore some further consequences of the choice of $\eps$.

\begin{itemize}
\item[(i)] If $\eps\simeq \eps_0>0$ then the system reduce to a simple Markov chain.\\[2mm]

\item[(ii)] If  $\eps\simeq \de$ then, as $\de,\tau\ra 0$ the system reduces to a drift plus a "fast" Markov chain. In fact

\[
\Lop^*\, = \, \left( \begin {array}{cc}
 (v_0-u_0)\pa_x( \cdot ) & 0\\
 0 & (v_1-u_1)\pa_x( \cdot )
 \end {array} \right)
\mbox{ and }\Kd^T\, = \frac{1}{\eps} \left( \begin {array}{cc} 
-k^0(x) & k^1(x)
\\\noalign{\medskip}k^0(x)& -k^1(x)
\end {array} \right), 
\]

\item[(ii)] If  $\eps\simeq \de^2$ and  
\[\frac{(v_i(\de,\tau)-u_i(\de,\tau))\de}{\eps}\simeq V_i\mbox{ for $i=0,1$}\]
for some $V_i$ as $\de,\tau\ra 0$, then the system reduces to an effective diffusion plus a "fast" Markov chain. In fact $\Ld^*$ and $\Kd^T$ read

\[
\Lop^*\, = \, \left( \begin {array}{cc}
 V_0\pa_x( \cdot ) + (v_0+u_0)\pa_x^2( \cdot ) & 0\\
 0 & V_1\pa_x( \cdot ) + (v_1+u_1)\pa_x^2( \cdot )
 \end {array} \right)
\mbox{ and }\Kd^T\, = \frac{1}{\eps} \left( \begin {array}{cc} 
-k^0(x) & k^1(x)
\\\noalign{\medskip}k^0(x)& -k^1(x)
\end {array} \right), 
\]
\end{itemize}

In the adiabatic regime the average dynamics appears to be an effective diffusion. In fact the invariant measure of the Markov chain is 

\[\mu=\left(\frac{k^1(x)}{k^1(x)+k^0(x)},\frac{k^0(x)}{k^1(x)+k^0(x)}\right).\]

Equation (\ref{average-dynamics}) then becomes

\begin{equation}
\begin{array}{ll}
\displaystyle\frac{\pa f^{(0)}(x,t)}{\pa t}=\frac{\pa}{\pa x}\left[\left(\frac{V_0k^1(x)}{k^1(x)+k^0(x)}+\frac{V_1k^0(x)}{k^1(x)+k^0(x)}\right)f^{(0)}(x,t)\right]+\\[4mm]
\displaystyle+
\frac{\pa^2}{\pa x^2}\left[\left(\frac{(u_0+v_0)k^1(x)}{k^1(x)+k^0(x)}+\frac{(u_1+v_1)k^0(x)}{k^1(x)+k^0(x)}\right)f^{(0)}(x,t)\right]
\end{array}
\end{equation}

\brem
Before introducing the switch reactions we make a comment about the continuum limit in the case in which
 a term
 
\[\int_0^t\der t'\frac{1}{\tau}\Kd_{\sig\sig'}[\vn,\de,t-t']P(\vn,\sig',t')\]

is present. One can observe that in this case the continuum limit depends on how the time scale $\tau$ is related with the scale at which the waiting time is defined. Possibly  there might be regimes where if $\tau$ is small enough.  Then

\[\int_0^t\der t'\frac{1}{\tau}\Kd_{\sig\sig'}[\vn,\de,t-t']P(\vn,\sig',t')\app\int_0^t\der t'\frac{1}{\tau}\Kop_{\sig\sig'}[\vx,t-t']\rho_{\sig'}(\vx,t'),\] 

where $\Kop_{\sig\sig'}[\vx,t-t']$ a new operator. The main problem is to identify some general minimal properties for such classes of scaling. 
\erem
 
\subsection{Switch reactions}
We now consider a set  of reactions that form an elementary "switch". This is essentially a system formed by two types of particles  (two chemical species) $A$ and $M$ interacting with a two-state system $\Sigma=\{O_0,O_1\}$. Particle $A$ regulates the switching and the two-state system 
which in turn regulates $M$. First we consider  a single switch. Its defining reactions are:

\[\begin{array}{ll}
A+O_0 \rlha[k^0(\de,\tau)]{k^1(\de,\tau)} O_1.
\end{array}\]

We show that if the dynamics of $A$ is included in the Master equation  (ME),
 the continuum limit and the adiabatic theory (the time scale of the process involving the finite states $\{O_0,O_1\}$) imply that 
 the noise is not Gaussian at order $O(\eps)$. This is related to the fact the the operator $\Ld^*$ in the ME is not diagonal, because there are reactions involving transition in $\LL$ and $\Sig$. Let us  consider the following two systems of reactions: 

\[
\begin{array}{llll}
\mbox{System n.1}\\[2mm]
A+O_0 \rlha[k^0(\de,\tau)]{k^1(\de,\tau)} O_1,\\[4mm]
O_1\rightarrow^{\nu(\de,\de)}\,O_0+M+A,\\[3mm]
M\ra^{\gamma(\de,\tau)}\emptyset.
\end{array}
\qquad\qquad
\begin{array}{lllll}
\mbox{System n.2}\\[2mm]
A+O_0 \rlha[k^0(\de,\tau)]{k^1(\de,\tau)} O_1,\\[4mm]
O_1\rightarrow^{\nu(\de,\tau)}\,O_1+M,\\[3mm]
M\ra^{\gamma(\de,\tau)}\emptyset.
\end{array}\]

The dynamics of the two reaction systems can be expanded asymptotically in $\eps$.
It turns out that the systems have the same deterministic limit but with different noise terms. Namely  
the dynamics n.1 and n.2 have same $O(1)$ (average dynamics) but 
differ from order $O(\eps)$.
We further analyse this point in System n.1 by including the $A$ dynamics. 

 \subsection{A more detailed analysis of a switch reaction}
 \label{switch}
 
 Consider the reactions
 
 \begin{equation}
 A+O_0 \rlha[k^0(\de,\tau)]{k^1(\de,\tau)} O_1.
\label{A-reaction}
\end{equation}

\noi We are interested in describing the reactions without assuming that $A$ particles are 
constant. We assume that there is a pool of $A$ from which particles are "created" and "annihilated". The annihilation from the pool corresponds to the absorption of an $A$ particle by $O_0$ and the transition to $O_1$. The creation corresponds to the releasing 
of an $A$ particle  from $O_1$ and the transition to $O_0$. In order to simplify the notation it is useful to introduce the following operators.  The state is identified by

\[P(a,t)=(P_0(a,t),P_1(a,t)).\]

Using reaction (\ref{A-reaction}) its ME reads

\begin{equation}
\left\{\begin{array}{ll}
\displaystyle\frac{\pa P_0(a,t)}{\pa t}=-(k^1(\de,\tau)/\tau)\,a\,P_0(a,t)+(k^0(\de,\tau)/\tau)P_1(a-1,t)\\[4mm]
\displaystyle\frac{\pa P_1(a,t)}{\pa t}=(k^1(\de,\tau)/\tau)\,(a+1)\,P_0(a+1,t)-(k^0(\de,\tau)/\tau)P_1(a,t)
\end{array}\right.
\end{equation}

\noi Now using the definition of $\Dop_a^\pm$ the ME can be rewritten as follows

\begin{equation}
\left\{\begin{array}{lllll}
\displaystyle\frac{\pa P_0(a,t)}{\pa t}=\overbrace{-(k^1(\de,\tau)/\tau)\,a\,P_0(a,t)+(k^0(\de,\tau)/\tau)P_1(a,t)}^{\mbox{Markov chain}}+\overbrace{\Dop_a^-((k^1(\de,\tau)/\tau)\,P_1(a,t))}^{\mbox{one $a$ is released}}\\[4mm]
\displaystyle\frac{\pa P_1(a,t)}{\pa t}=\overbrace{(k^1(\de,\tau)/\tau)\,a\,P_0(a,t)-(k^0(\de,\tau)/\de)P_1(a,t)}^{\mbox{Markov chain}}+\overbrace{\Dop_a^+((k^0(\de,\tau)/\tau)\,a\,P_1(a,t)).}^{\mbox{one $a$ is absorbed}}
\end{array}\right.
\end{equation}

\noi The ME can be recast as in (\ref{gen-ME1-ad}) by defining

\[
\Ld^*[a,\de,\tau]\, = \, \left( \begin {array}{cc}
 0 &\Dop_a^-((k^1(\de,\tau)/\tau)( \cdot ))
 \\\noalign{\medskip}\Dop_a^+((k^0(\de,\tau)/\tau)\,a\,( \cdot )) & 0
 \end {array} \right) ,
\]

\noi and

 \[
\Kd^T[a,\de,\tau]\, =  \left( \begin {array}{cc} 
-a\,k^0(\de,\tau)/\tau & k^1(\de,\tau)/\tau
\\\noalign{\medskip}a\,k^0(\de,\tau)/\tau & -k^1(\de,\tau)/\tau
\end {array} \right).
\]

\brem
$\Ld^*$ is non-diagonal and depends on the Markov chain parameters. 
For the adiabatic limit the continuum limit is needed. The operators $\Dop^-((k^1(\de,\tau)/\tau)( \cdot ))$ and $\Dop^+((k^0(\de,\tau)/\tau)\,a\,( \cdot ))$ can have a finite limit 
as $\de,\tau\ra 0$. 
\erem

\subsection{Continuum limit}
Let us assume that for $\de,\tau\ra 0$ we take 

\[\frac{k^0(\de,\tau)}{\de\,\tau}\simeq \frac{1}{\eps}\,{k^0},~~\frac{k^1(\de,\tau)}{\tau}\simeq \frac{1}{\eps}\,{k^1}.\]

for some $\eps=\eps(\de,\tau)$ and $k^0,k^1>0$. Then the difference operators will have the following asymptotic exapnsions:

\[\Dop_a^-((k^1(\de,\tau)/\tau)\,( \cdot ))\app-\frac{k^1\,\de}{\eps}\frac{\pa ( \cdot )}{\pa x}+\frac{k^1\,\de^2}{\eps}\,\frac{\pa^2( \cdot )}{\pa x^2},\]

and

\[\Dop_a^+((k^0(\de,\tau)/\tau)\,a\,( \cdot ) )\app\frac{k^0\,\de}{\eps}\frac{\pa (x( \cdot ))}{\pa x}+\frac{k^0\,\de^2}{\eps}\,\frac{\pa^2(x\,( \cdot ))}{\pa x^2}\]

\noi Taking $\tau,\de\ra 0$ such that

\[ \frac{1}{\tau}\,o(\de^2)\ra 0, \] 

the limit of the ME reads

\[\frac{\pa \rho}{\pa t}=\Lop^*\rho+\frac{1}{\eps}\,\Kd^T\rho.\]

The form of the operators $\Lop$ and $\Kop$ is identified by the following two
cases: 

\begin{itemize}
\item[(i)] if $\eps(\de,\tau)\simeq \eps_0$ as $\de,\tau\ra 0$ then
\[\Lop=0\mbox{ and }\Kd^T\, =  \left( \begin {array}{cc} 
-x\,k^0 & k^1
\\\noalign{\medskip}x\,k^0& -k^1
\end {array} \right), 
\]

\item[(ii)] if $\eps(\de,\tau)\simeq\de$ as $\de,\tau\ra 0$ then
\[
\Lop^*\, = \, \left( \begin {array}{cc}
 0 &-\pa_x(\,k^1\,( \cdot ))
 \\\noalign{\medskip}\pa_x(k^0\, x\,( \cdot )) & 0
 \end {array} \right)
\mbox{ and }\Kd^T\, =  \left( \begin {array}{cc} 
-x\,k^0 & k^1
\\\noalign{\medskip}x\,k^0& -k^1
\end {array} \right).
\]
\end{itemize}

\noi We like to consider the case (ii). In this condition we can apply 
adiabatic theory. First note that $\Kd^T$ invariant measure is

\[\mu=\left(\frac{k^0}{k^0x+k^1},\frac{k^1x}{k^0x+k^1}\right).\]

\noi  The adiabatic limit can be computed. In particular at order $O(1)$ we found

\[\frac{\pa f^{(0)}}{\pa t}=\langle\be,\Lop^*(\mu f^{(0)})\rangle=\frac{\pa}{\pa x}((-k^1\mu_1+k^0x\,\mu_0)f^{(0)}).\]

\noi Using the explicit form of the invariant measure it is easy to verify that
\[ (-k^1\mu_1+k^0x\,\mu_0)\]
identically vanish, which means that the concentration $x=[A]$ is constant along the average dynamics.

\subsubsection{Noise at order $O(\eps)$}
As shown in \cite{LM1}, at order $O(\eps)$ the noise can be evaluated by computing

\[\langle\be,\Lop^*(\xi^{(1)})\rangle=-\langle\be,\Lop^*\,(\Kd^T)^D\,\Lop^*(\mu f^{(0)})\rangle.\]

Now

\[(\Kd^T)^D\,\Lop^*(\mu f^{(0)})=\frac{1}{(k^0\,x+k^1)^2}
\left(\begin{array}{cc}
-k^0\,x & k^1\\
k^0\,x & -k^1
\end{array}\right)\,
\left(\begin{array}{c}
-\pa_x(k^1\,\mu_1\,f^{(0)})\\
\pa_x(k^0\,x\,\mu_0\,f^{(0)})
\end{array}\right)\]
which is equal to
\[(\Kd^T)^D\,\Lop^*(\mu f^{(0)})=\frac{1}{(k^0\,a+k^1)^2}
\left(\begin{array}{c}
k^0\,x\,\pa_x(k^1\,\mu_1\,f)+k^1\,\pa_x(k^0\,x\,\mu_0\,f^{(0)})\\
-k^0\,x\,\pa_x(k^1\,\mu_1\,f)-k^1\,\pa_x(k^0\,x\,\mu_0\,f^{(0)})
\end{array}\right).\]

Using $\mu_0+\mu_1=1$ the expression

\[k^0\,x\,\pa_x(k^1\,\mu_1\,f^{(0)})+k^1\,\pa_x(k^0\,x\,\mu_0\,f^{(0)})\]

can be rewritten as

\[k^0\,k^1\,x\,\pa_xf^{(0)}+k^0\,k^1\,\mu_0\,f^{(0)}, \]

so

\[(\Kd^T)^D\,\Lop^*(\mu f^{(0)})=\frac{1}{(k^0\,a+k^1)^2}
\left(\begin{array}{c}
k^0\,k^1\,x\,\pa_xf^{(0)}+k^0\,k^1\,\mu_0\,f^{(0)}\\
-k^0\,k^1\,x\,\pa_xf^{(0)}-k^0\,k^1\,\mu_0\,f^{(0)}.
\end{array}\right).\]

Finally the noise term can be computed. It is equal to

\[\begin{array}{ll}
\displaystyle-\langle\be,\Lop^*\,(\Kd^T)^D\,\Lop^*(\mu f^{(0)})\rangle=
-\frac{\pa}{\pa x}\left(\frac{k^0\,(k^1)^2\,x}{(k^0\,x+k^1)^2}\,\frac{\pa f^{(0)}}{\pa x}+
\frac{k^0\,(k^1)^2\,\mu_0}{(k^0\,x+k^1)^2}\,f^{(0)}+\right.\\[5mm]
\displaystyle\left.+\frac{(k^0\,x)^2\,k^1}{(k^0\,x+k^1)^2}\,\frac{\pa f^{(0)}}{\pa x}+
\frac{(k^0)^2\,x\,k^1\,\mu_0}{(k^0\,x+k^1)^2}\,f^{(0)}\right).
\end{array}\]

\noi It is not difficult to show that the noise term determines an 
elliptic operator which is negative definite. We can therefore conclude 
that the noise at order $O(\eps)$ does not determine a genuine Fokker-Planck 
equation and therefore the time evolution of the concentration $x$ cannot be described - on short time scales (see \cite{Kurtz1}, \cite{Kurtz2}, \cite{Kurtz3}) - through an Ito stochastic differential equation.

\subsection{Analisys of systems n.1 and n.2}
Let us consider reactions in system n.1:

\[\begin{array}{llll}
A+O_0 \rightarrow^{k^0(\de,\tau)} O_1,\\[3mm]
O_1 \rightarrow^{k^1(\de,\tau)} O_0+A,\\[3mm]
O_1\rightarrow^{\nu(\de,\tau)}\,O_0+M+A,\\[3mm]
M\ra^{\gamma(\de,\tau)}\emptyset.
\end{array}\]

The state of the system is defined by the probabilities

\[P(m,a,t)=(P_0(m,a,t),P_1(m,a,t)).\]

\brem
Without loss of generality we can assume 
that the number of $A$ particles $a$ is considered large: $a\pm 1\simeq a$. 
\erem

The ME reads

\begin{equation}
\left\{\begin{array}{lllll}
\displaystyle\frac{\pa P_0(m,a,t)}{\pa t}=\overbrace{-a\,(k^1(\de,\tau)/\tau)P_0(m,a,t)+(k^0(\de,\tau)/\tau)P_1(m,a,t)}^{\mbox{Markov chain}}+\\[4mm]
\displaystyle\overbrace{-(\gamma(\de,\tau)/\tau)\,m\,P_0(m,a,t)+(\gamma(\de,\tau)/\tau\,(m+1)\,P_0(m+1,a,t)}^{\mbox{degradation}}+\\[4mm]
\displaystyle\overbrace{+(\nu(\de,\tau)/\tau)\,P_1(m-1,a,t)}^{\mbox{creation of a $m$ and transition to $O_0$}} \\[4mm]
\displaystyle\frac{\pa P_1(m,a,t)}{\pa t}=\overbrace{a\,(k^1(\de,\tau)/\tau)P_0(m,a,t)-(k^0(\de,\tau)/\tau)P_1(m,a,t)+}^{\mbox{Markov Chain}}\\[4mm]
\displaystyle\overbrace{-(\gamma(\de,\tau)/\tau)\,m\,P_1(m,a,t)+(\gamma(\de,\tau)/\tau)\,(m+1)\,P_1(m+1,a,t)}^{\mbox{degradation}}+\\[4mm]
\displaystyle\overbrace{-(\nu(\de,\tau)/\tau)\,P_1(m,a,t)}^{\mbox{creation of a $m$ and transition to $O_0$}} 
\end{array}\right.
\end{equation}

\noi In matrix form the ME reads

\[\frac{\pa P(m,a,t)}{\pa t}=\Ld^*\,P(m,a,t)+\Kd^T\,P(m,a,t), \]

\noi where $P(m,a,t)=(P_0(m,a,t),P_1(m,a,t))$ and the operator $\Ld^*$ is

\[
\Ld^*\, = \, \frac{1}{\tau}\left( \begin {array}{cc}
 \gamma(\de,\tau)\,\Dop^+_m(m\,( \cdot ) \,) &\nu(\de,\tau)\,\Eop^-_m
 \\\noalign{\medskip}0 &  \gamma(\de,\tau)\,\Dop^+_m(m\,( \cdot ) \,)-\nu(\de,\tau)\,\id
 \end {array} \right).
\]

\brem
Note that the matrix of the operator $\Ld^*$ is not diagonal, but the theory developed in \cite{LM1}
still applies.
\erem

\noi The Markov chain has transpose generator given by

 \[
\Kd^T\, =  \frac{1}{\tau}\left( \begin {array}{cc} 
-a\,k^0(\de,\tau) & k^1(\de,\tau)
\\\noalign{\medskip}a\,k^0(\de,\tau)& -k^1(\de,\tau)
\end {array} \right).
\]

\noi Its invariant measure is

\[\mu=\left(\frac{k^1(\de,\tau)}{k^0(\de,\tau)\,a+k^1(\de,\tau)},\frac{k^0(\de,\tau)\,a}{k^0(\de,\tau)\,a+k^1(\de,\tau)}\right).\]

\begin{assumption}[Adiabatic assumption]
We now assume that without performing the scaling $\de\ra 0$ and $\tau\ra 0$ the 
time on which the Markov chain on $\Sig$ reaches its equilibrium measure is faster than 
the time evolution of $m,a$. Therefore we can make the following formal substitution 
\[\Kop\ra\frac{1}{\eps}\,\Kop.\]
\end{assumption}

At this state we can construct the solution by the asymptotic expansion in $\eps$ according to the scheme
 developed in \cite{LM1}.

\subsubsection*{Average dynamics}
The leading order term of the expansion is given by

\[\frac{\pa f^{(0)}(\bn,t)}{\pa t}=\langle\be_\mu,\Ld^*(\mu(\bn)\,f^{(0)}(\bn,t))\rangle,\]

This is the average dynamics. In the present example note that $\be_\mu=\be=(1,1)$ for there is 
a unique invariant measure. Using the expression of $\Ld^*$. An explicit calculations yield 

\[\begin{array}{ll}
\displaystyle\langle\be_\mu,\Ld^*(\mu(\bn)\,f^{(0)}(\bn,t))\rangle=\\[3mm]
\displaystyle=(\gamma(\de,\tau)/\tau)\Dop^+_m(\mu_0\,m\,f^{(0)}(m,a,t))
+(\gamma(\de,\tau)/\tau)\Dop^+_m(\mu_1\,m\,f^{(0)}(m,a,t))+\\[3mm]
\displaystyle +(\nu(\de,\tau)/\tau)\,\Eop^-(\mu_1\,f^{(0)}(m,a,t))-(\nu(\de,\tau)/\tau)\,\mu_1\,f^{(0)}(m,a,t).
\end{array}\]

\noi Now use $\mu_0+\mu_1=1$ and the explicit expression of $\mu_1$ to obtain
 
\[\begin{split}
\displaystyle\langle\be_\mu,\Ld^*(\mu(\bn)\,f^{(0)}(\bn,t))\rangle=
(\gamma(\de,\tau)/\tau)\Dop^+_m(m\,f^{(0)}(m,a,t))+&\\[4mm]
\displaystyle+(\nu(\de,\tau)/\tau)\,\Dop^-_m\left(\frac{k^0(\de,\tau)\,a}{k^0(\de,\tau)\,a+k^1(\de,\tau)}f^{(0)}(m,a,t)\right).&
\end{split}
\]

\noi Therefore the average dynamics is given by the following Master equation

\begin{equation}
\label{e:average-dyn}
\begin{split}
\displaystyle\frac{\partial f^{(0)}(m,a,t)}{\partial t}=(\gamma(\de,\tau)/\tau)\Dop^+_m(m\,f^{(0)}(m,a,t))+&\\[4mm]
\displaystyle+(\nu(\de,\tau)/\tau)\,\Dop^-_m\left(\frac{k^0(\de,\tau)\,a}{k^0(\de,\tau)\,a+k^1(\de,\tau)}f^{(0)}(m,a,t)\right).
\end{split}
\end{equation}

\noi An important observation is the following:

\brem
Note that by taking a reaction (like in system n.2) 

\begin{equation}
O_1\rightarrow^{\nu(\de,\tau)}\,O_1+M
\label{r2}
\end{equation}

\noi instead of 

\begin{equation}
O_1\rightarrow^{\nu(\de,\tau)}\,O_0+M+A,
\label{r1}
\end{equation}

we would obtain the an average dynamics equal to equation (\ref{e:average-dyn}). 
\erem

\noi From the previous remark one can show that the continuum 
limit of the average dynamics for a system containing reaction (\ref{r2}) coincides with 
the one containing reaction (\ref{r1}). Let us observe that the difference in the two systems of reactions appear only in the operator $\Ld^*$. For systems with reaction (\ref{r2})  the operator $\Ld^*$ is a diagonal matrix with entries difference operators. For  a system with reaction (\ref{r1})  the operator $\Ld^*$ is a no longer a diagonal matrix. In order to  see the difference between the two systems it is necessary to look at higher order terms in the expansion. The generic order $O(\eps)$ corrections are

\begin{equation}
\left\{\begin{array}{ll}
\displaystyle\xi^{(1)}(\bn,t)=-(\Kd^T_\mu)^D\Ld^*(\mu(\bn)\,f^{(0)}(\bn,t))\\[4mm]
\displaystyle\frac{\pa f^{(1)}(\bn,t)}{\pa t}=\langle\be_\mu,\Ld^*(\mu(\bn)\,f^{(1)}(\bn,t))\rangle+\langle\be_\mu,\Ld^*(\xi^{(1)}(\bn,t))\rangle.
\end{array}
\right.
\label{dual-step1}
\end{equation}

\noi Consider the two systems of reactions
\[
\begin{array}{llll}
\mbox{System n.1}\\[2mm]
A+O_0 \rightarrow^{k^0(\de,\tau)} O_1,\\[3mm]
O_1 \rightarrow^{k^1(\de,\tau)} O_0+A,\\[3mm]
O_1\rightarrow^{\nu(\de,\tau)}\,O_0+M+A,\\[3mm]
M\ra^{\gamma(\de,\tau)}\emptyset.
\end{array}
\qquad\qquad
\begin{array}{lllll}
\mbox{System n.2}\\[2mm]
A+O_0 \rightarrow^{k^0(\de,\tau)} O_1,\\[3mm]
O_1 \rightarrow^{k^1(\de,\tau)} O_0+A,\\[3mm]
O_1\rightarrow^{\nu(\de,\tau)}\,O_1+M,\\[3mm]
M\ra^{\gamma(\de,\tau)}\emptyset.
\end{array}\]

\noi These systems differ only in the form of the operator $\Ld^*$. For system n.1

\[
\Ld_1^*\, = \, \frac{1}{\tau}\left( \begin {array}{cc}
 \gamma(\de,\tau)\,\Dop^+_m(m\,( \cdot ) \,) &\nu(\de,\tau)\,\Eop_m^-(( \cdot ))
 \\\noalign{\medskip}0 &  \gamma(\de,\tau)\,\Dop^+_m(m\,( \cdot ) \,)-\nu(\de,\tau)\,\id
 \end {array} \right),
\]

\noi and for system n.2

\[
\Ld_2^*\, = \, \frac{1}{\tau}\left( \begin {array}{cc}
 \gamma(\de,\tau)\,\Dop^+_m(m\,( \cdot ) \,) &0
 \\\noalign{\medskip}0 &  \gamma(\de,\tau)\,\Dop^+_m(m\,( \cdot ) \,)-\nu(\de,\tau)\,\Dop^-_m(( \cdot ))
 \end {array} \right).
\]

\noi The two systems have the same Markov chain therefore same invariant measure, and  the same
Drazin inverse $(\Kd_\mu^T)^D$. We have already pointed out  that the average dynamics is the same for both systems, but now let us look at the first order corrections in (\ref{dual-step1}).  Both systems have the same solution $f^{(0)}(\bn,t)$ to the average equation. It is sufficient to consider the equation for $\xi^{(1)}(\bn,t)$. In fact for system n.1 

\[\xi_1^{(1)}(\bn,t)=-(\Kd^T_\mu)^D\Ld_1^*(\mu(\bn)\,f^{(0)}(\bn,t)),\]

\noi and for system n.2

\[\xi_2^{(1)}(\bn,t)=-(\Kd^T_\mu)^D\Ld_2^*(\mu(\bn)\,f^{(0)}(\bn,t)).\]

\noi Now take the difference of the equations to obtain 

\[\xi_1^{(1)}(\bn,t)-\xi_2^{(1)}(\bn,t)=-(\Kd^T_\mu)^D\Ld_1^*(\mu(\bn)\,f^{(0)}(\bn,t))
+(\Kd^T_\mu)^+\Ld_2^*(\mu(\bn)\,f^{(0)}(\bn,t)).\]

\noi We then use the linearity of the operators to write

\[\xi_1^{(1)}(\bn,t)-\xi_2^{(1)}(\bn,t)=(\Kd^T_\mu)^D(\Ld_2^*-\Ld^*_1)(\mu(\bn)\,f^{(0)}(\bn,t)).\]

\noi Now using the explicit expression of $\Ld^*$ operators, one finds:

\[\Ld^*_2-\Ld_1^*=\frac{1}{\tau}\left( \begin {array}{cc}
 0 &-\nu(\de,\tau)\,\Eop^-_m(( \cdot ))
 \\\noalign{\medskip}0 &  -\nu(\de,\tau)\,(\Dop^-_m-\id)(( \cdot ))
 \end {array} \right).\]

\noi The difference $\Ld_1^*-\Ld_2^*$ is not identically zero, therefore we can conclude that \emph{ system n.1 and n.2 have adiabatic limits which generate two stochastic processes which differ at order $O(\eps)$.}\\
Now suppose that we take the continuum limit. Clearly it can be shown (from \cite{LM1}) that up to order $O(\eps)$ the dynamics is described by a stochastic differential equation (SDE), whenever 
$\Ld^*$ is diagonal. The noise term is essentially related to the formula for $\xi^{(1)}(\bn,t)$, which contains at order $O(\eps)$ the differences between system n.1 and n.2.
 
\subsubsection{Explicit construction of the noise} 
We now compute the noise term (up to order $O(\eps)$) for the systems n.1 and n.2. First recall that in both cases

\[(\Kd_\mu^T)^D= \, \frac{\tau}{(a \,k^0(\de,\tau)+k^1(\de,\tau))^2}\left( \begin {array}{cc} -a \,k^0(\de,\tau)& k^1(\de,\tau)\\\noalign{\medskip} a \,k^0(\de,\tau)& -k^1(\de,\tau) \end {array} \right). 
\]

 \subsubsection*{System n.1}
The noise is 
\[\langle\be,\Ld_1^*(\xi^{(1)})\rangle=-\langle\be,\Ld_1^*\,(\Kd^T)^D\,\Ld_1^*(\mu f^{(0)})\rangle.\]

\noi After some lengthy but simple calculations one finds

\[\begin{array}{ll}
\displaystyle-\langle\be,\Ld_1^*\,(\Kd^T)^D\,\Ld_1^*(\mu f^{(0)})\rangle=\\[3mm]
\displaystyle=\frac{-\tau}{(a \,k^0+k^1)^2}\left[
(-k^1\mu_1\nu\gamma+ak^1\mu_0\nu\gamma)\Dop^-_m(\Dop^+_m(mf^{(0)}))\right.\\[3mm]
\displaystyle\left.+ak^0\mu_1\nu^2\Dop^-_m(\Eop^-_m(f^{(0)})+k^1\mu_1\nu^2\Dop^-_m(f^{(0)})\right]
\end{array}
\]

\noi It is not hard to see that the noise term gives rise to a parabolic operator which is 
not always positive definite.

\subsubsection*{System n.2}
For system n.2 performing the same calculations one finds 
\[\begin{array}{ll}
\displaystyle-\langle\be,\Ld_2^*\,(\Kd^T)^D\,\Ld_2^*(\mu f^{(0)})\rangle=
\frac{2ak^0k^1\gamma^2\,\tau}{(a \,k^0+k^1)^3}
\Dop^+_m(m\Dop^+_m(mf^{(0)}))+\\[3mm]
\displaystyle +\frac{ak^0k^1\nu\gamma\,\tau}{(a \,k^0+k^1)^3}
[\Dop^-_m(\Dop^+_m(mf^{(0)}))+\Dop^+_m(m\Dop^-_m(f^{(0)}))]+
\frac{ak^0k^1\nu^2\,\tau}{(a \,k^0+k^-)^3}(\Dop_m^-(\Dop^-_m
(f^{(0)})))
\end{array}
\]
\brem
For system n.2 it is crucial that $\Ld_2^*$ is diagonal.
\erem

\section{Detailed analysis of system n.1}
In this section system n.1 is analysed assuming that also $A$ molecules 
have a dynamics. Now the ME reads

\begin{equation}
\left\{\begin{array}{llll}
\displaystyle\frac{\pa P_0(m,a,t)}{\pa t}=-(k^1(\de,\tau)/\tau)\,a\,P_0(m,a,t)+(k^0(\de,\tau)/\tau)P_1(m,a-1,t)+\\[3mm]
\displaystyle+(\nu(\de,\tau)/\tau)\,P_1(m-1,a-1,t)+\Dop^+_m((\gamma(\de,\tau)/\tau)\, m\, P_0(m,a,t))\\[4mm]
\displaystyle\frac{\pa P_1(m,a,t)}{\pa t}=(k^1(\de,\tau)/\tau)\,(a+1)\,P_0(m,a+1,t)-(k^0(\de,\tau)/\tau)P_1(m,a,t)+\\[3mm]
\displaystyle-(\nu(\de,\tau)/\tau)\,P_1(m,a)+D_m^+((\gamma(\de,\tau)/\tau)\,m\,P_1(m,a,t)).
\end{array}\right.
\end{equation}

\noi The ME is not in the form of an operator plus a Markov chain. To rewrite it in that form one can use the operators $\Dop^\pm$ and $\Eop^\pm$:

\begin{equation}
\left\{\begin{array}{lllll}
\displaystyle\frac{\pa P_0(m,a,t)}{\pa t}=-(k^1(\de,\tau)/\tau)\,a\,P_0(m,a,t)+[(k^0(\de,\tau)/\tau)+(\nu(\de,\tau)/\tau)]P_1(m,a,t)+\\[4mm]
\displaystyle+\Dop^-_a([(\nu(\de,\tau)/\tau)+(k^1(\de,\tau)/\tau)]\, P_1(a,m,t))+\Dop_m^+((\gamma(\de,\tau)\tau)\,m\, P_0(m,a,t))+\\[4mm]
\displaystyle+\Dop^-_m((\nu(\de,\tau)/\tau)\,\Eop_a^-(P_1(m,a,t)))\\[5mm]

\displaystyle\frac{\pa P_1(m,a,t)}{\pa t}=(k^1(\de,\tau)/\tau)\,a\,P_0(m,a,t)-[(k^0(\de,\tau)/\tau)+(\nu(\de,\tau)/\de)]P_1(a,t)+\\[4mm]
\displaystyle+\Dop^+_a((k^0(\de,\tau)/\tau)P_0(m,a,t))+\Dop_m^-((\gamma(\de,\tau)/\tau)\,m\,P_1(m,a,t))
\end{array}\right.
\end{equation}

\noi Now the form (\ref{gen-ME1-ad}) is obtained by taking

 \[
\Kd^T\, =  \frac{1}{\tau}\left( \begin {array}{cc} 
-a\,k^0(\de,\tau) & k^1(\de,\tau)+\nu(\de,\tau)
\\\noalign{\medskip}a\,k^0(\de,\tau)& -k^1(\de,\tau)-\nu(\de,\tau)
\end {array} \right), 
\]

and

\[
\Ld^*\, = \, \frac{1}{\tau}\left( \begin {array}{cc}
 \Dop^+_m(\gamma(\de,\tau)\,m\,( \cdot ) \,) &\Dop^-_a([\nu(\de,\tau)+k^1(\de,\tau)]\,( \cdot ))+\Dop_m^-(\nu(\de,\tau)\,\Eop^-_a(( \cdot )))
 \\\noalign{\medskip}\Dop_a^+(a\,k^0(\de,\tau)\,( \cdot )) &  \Dop^+_m(\gamma(\de,\tau)\,m\,( \cdot ) \,)
 \end {array} \right) ,
\]

\subsection{Continuum limit}
Looking at the terms in the operators $\Kd^T$ and $\Ld^*$ 
it is possible to guess a limit behaviour. We chose a scaling as follows

\[\frac{k^0(\de,\tau)}{\de\,\tau}\simeq\frac{k^0}{\eps},~~\frac{\de\,k^1(\de,\tau)}{\tau}\simeq\frac{k^1}{\eps},
~~\frac{\nu(\de,\tau)}{\tau}\simeq\frac{\nu}{\eps},~~\frac{\ga(\de,\tau)}{\de\,\tau}\simeq\frac{\ga}{\eps}.\]

\noi Then proceeding as in the examples above we take $\eps\simeq\de$ as $\de,\eps\ra 0$ and we obtain
 \[
\Kd^T\app \Kd^T = \frac{1}{\eps} \left( \begin {array}{cc} 
-a\,k^0 & k^1+\nu
\\\noalign{\medskip}a\,k^0& -k^1-\nu
\end {array} \right), 
\]

and

\[
\Ld^*\app \Lop^* = \, \left( \begin {array}{cc}
 \pa_m(\gamma\,m\,( \cdot ) \,) &-\pa_a([\nu+k^1]\,( \cdot ))-\pa_m(\nu(( \cdot )))
 \\\noalign{\medskip}\pa_a(a\,k^0\,( \cdot )) &  \pa_m(\gamma\,m\,( \cdot ) \,)
 \end {array} \right) ,
\]

\noi Then the limit ME have the form

\begin{equation}
\frac{\pa\rho(m,a,t)}{\pa t}=\Lop^*(\rho(m,a,t))+\frac{1}{\eps}\,\Kd^T\rho(m,a,t)
\label{limitME}
\end{equation}

\brem
Note that also in this case there is no diffusion term in $\Lop^*$.
\erem

\noi Upon the validity of (\ref{limitME}) the theory in \cite{LM1} applies and 
an average dynamics can be computed. The new invariant measure is
\[\mu=\left(\frac{k^1+\nu}{k^0\,a+k^1+\nu},\frac{k^0\,a}{k^0\,a+k^1+\nu}\right).\]

\noi The ME at order $O(\eps^0)$ is

\[\frac{\pa f^{(0)}}{\pa t}=\langle \be,\Lop^*(\mu\,f^{(0)})\rangle, \]

which turns out to be

\[\frac{\pa f^{(0)}}{\pa t}=\frac{\pa}{\pa m}[(\gamma\,m\,(\mu_1+\mu_0)-\nu\,\mu_1)f^{(0)}]+\frac{\pa}{\pa a}[-(k^1+\nu)\mu_1+\mu_0\,k^0\,a)f^{(0)}].\]

This corresponds to an average dynamics being equal to:

\[\left\{\begin{array}{ll}
\displaystyle \dot{m}(t)=-\gamma\,m(t)+\frac{\nu\,k^0\,a(t)}{k^0\,a(t)+k^1+\nu},\\[4mm]
\displaystyle \dot a(t)=0.
\end{array}\right. \]

\noi As in the switch reaction (section \ref{switch}) the average dynamics of $a$ is trivial. Let $a(0)$ the initial value of $A$ molecules. Then the steady state value for $M$ molecules is

\[\overline{m}=\frac{\nu\,k^0\,a(0)}{\gamma\,(k^0\,a(0)+k^1+\nu)}.\]

 
\section{Conclusions}
This paper shows a possible way to model reactions networks containing possibly more than standard mass-action kinetics. We used finite states which can model conformational changes of larger molecules, like proteins and corresponding binding/unbinding events
of smaller molecules, typically substrates binding to the protein. The finite number of such larger proteins can be included in the structure of the finite state Markov chain, as it has been illustrated in \cite{LM1} and \cite{LM2}. The approach is based on the analysis of the different scales (spatio-temporal and number of particles) present in the system. The multiscale analysis leads naturally to the idea to compare the dynamics of the discrete states with the dynamics governed by  mass-action  kinetics. This comparison is done through adiabatic theory. The theory has many applications, primarily in cell biology to describe both the spatial and spatially averaged dynamics of macro-molecular interaction with substrates, ions or transcription factors. Obviously it can also be used in completely different branches of science where a similar setting can be defined in a meaningful way. Some simple examples  show this large potential of the approach. In future work the theory will be extended to systems whose underline stochastic dynamics is not Markovian. This is especially relevant for different cellular transport processes. As we could derive macroscopic equations from microscopic interactions, for example also capturing all classical enzyme kinetics, the next task in the analysis is to find ways of analysis how large systems of macroscopic reaction kinetics behave qualitatively, given their interaction is described by a graph containing the essential information of these interactions. Such approaches are reviewed in \cite{Mirela}.
 
\subsubsection*{Acknowledgements}
The authors would like to thank Mirela Domijan for thorough 
reading the paper and for her comments. 
This paper is part of the research activities supported by 
UniNet contract 12990  funded by the 
European Commission in the context of the VI Framework Programme.

\end{document}